\documentclass[12pt]{article}
\usepackage{amsmath,cite,a4wide}
\usepackage{slashed}
\usepackage{axodraw4j}
\usepackage{pstricks}
\usepackage{color}
\usepackage{changepage}
\usepackage{tabularx}
\usepackage{graphics}
\usepackage{array}
\usepackage[utf8]{inputenc}
\newcommand{\MSbar}{\overline{\rm{MS}}}
\def\nn{\nonumber\\}
\def\be{\begin{equation}}
\def\ee{\end{equation}}
\def\pa{\partial}

\def\tr{\hbox{tr}}
\def\tz{\tilde z}
\def\tzeta{\tilde \zeta}
\def\tZ{\tilde Z}
\def\tG{\tilde G}
\def\tx{\tilde x}
\def\txi{\tilde\xi}

\def\psibar{\overline \psi}

\def\yhat{\hat y}

\title{test}
\author{dij }
\date{March 2016}
\begin{document}
\begin{titlepage}
\begin{flushright}
LTH1080\\
\end{flushright}
\date{}
\vspace*{3mm}

\begin{center}
{\Huge Consistency of non-minimal renormalisation schemes}\\[12mm] {\bf I.~Jack\footnote{{\tt dij@liv.ac.uk}} and C.~Poole\footnote{{\tt c.poole@liv.ac.uk}}}\\ 

\vspace{5mm}
Dept. of Mathematical Sciences,
University of Liverpool, Liverpool L69 3BX, UK\\

\end{center}

\vspace{3mm}
\begin{abstract}
Non-minimal renormalisation schemes such as the momentum subtraction scheme (MOM) have frequently been used for physical computations. The consistency of such a scheme relies on the existence of a coupling redefinition linking it to 
$\MSbar$. We discuss  the implementation of this procedure in detail for a general theory and show how to construct the relevant redefinition up to three-loop order, for the case of a general theory of fermions and scalars in four dimensions and a general scalar theory in six dimensions. 
\end{abstract}

\vfill

\end{titlepage}
\numberwithin{equation}{section}

\section{Introduction}
It is expected that results obtained in one renormalisation scheme will be related to those obtained in another  by a (finite) redefinition of the couplings and fields of the theory. The required redefinition may be easily constructed by considering the finite differences between the bare quantities in the two schemes. In particular, the $\beta$-function in one scheme may be derived from the $\beta$-function in the other, in a well-known manner. The most widely-used scheme is of course minimal subtraction within dimensional regularisation, or more precisely its variant $\MSbar$. In this paper we discuss the issues involved in relating a general non-minimal scheme to $\MSbar$ in two cases:  a general theory of fermions and scalars with multiple fields and couplings in four dimensions, and a general scalar theory with multiple fields and couplings in six dimensions. The procedure is implicit in the standard literature\cite{collins} in general terms, but the detailed implementation for a theory of multiple fields and couplings presents some problems, especially in connection with one-particle reducible (1PR) 2-point function contributions, which we elucidate here. Our results apply to any non-minimal scheme; however at appropriate points we shall emphasise especially the application to the most widely-used non-minimal scheme, namely momentum subtraction (MOM).
The MOM scheme was first introduced in Ref.~\cite{jag5} for a three-loop computation in six-dimensional $\phi^3$ theory--one of the theories we shall consider in detail later. However it was used more widely following its independent development for QCD in Ref.~\cite{jag4}. The
renormalisation constants of the MOM scheme are defined such that all the Green's functions are set to their tree values at the
subtraction point\cite{jag4}. Variations (called ``hybrid'' MOM schemes) have also  been discussed where the renormalisation constants associated with $2$-point functions
are defined in a different way to the coupling-constant renormalisations\cite{smekal} (see Refs.~\cite{jag6,kataev} for further computations in this scheme). Motivated by this, we shall consider similar  ``hybrid'' schemes in the general non-minimal case. 

 We shall describe our results here in a little more detail, before presenting the full calculation in the following sections. The coupling redefinition which takes us from $\MSbar$ to a non-minimal scheme may readily be derived. When we implement it, we find a non-zero prediction at three-loop order for certain $\beta$-function terms corresponding to 1PR contributions to the anomalous dimension $\gamma$. More precisely, these terms depend on the antisymmetric part of a certain 1PR contribution to $\gamma$. One's natural expectation might be that such terms should vanish; but when one reviews the complications involved in performing an explicit computation, it becomes clear that intuition may not be a reliable guide.  For a non-minimal subtraction scheme such as MOM, the process of computing the $\beta$-function from the bare couplings of the theory is non-trivial, and an additional complexity is introduced when one recalls that the computation of the two-point function in fact only determines the symmetric part. We resort to an indirect argument to obtain the antisymmetric part, based upon working back from the anomalous dimension as obtained by scheme change from $\MSbar$.  Finally, it is important to recall that the coupling redefinition which effects the scheme change is accompanied by a corresponding field redefinition. This has no effect on the prediction for the $\beta$-function, but crucially it does affect the anomalous dimension and thereby our indirect method of computation of  the $\beta$-function in the non-minimal scheme.

The layout of the paper is as follows. We consider first a scalar-fermion theory in four dimensions in Sect. 2, and then a scalar theory in six dimensions in Sect. 3. The renormalisation of these theories is familiar in the literature, at least as far as 1PI contributions are concerned. We therefore try to present only as much background is needed to follow the arguments. We finish with conclusions in Sect. 4 and list some explicit diagrammatic results in Appendix A, and basic general results from renormalisation theory in Appendix B.

\section{Scalar-fermion theory in four dimensions}
We consider here a theory in four dimensions containing $n_{\psi}$ two-component Weyl fermion fields $\psi_a$, and $n_{\phi}$ real scalar fields $\phi^i$, $i=1,\ldots n_{\phi}$. The lagrangian is
\be
L=-\pa^{\mu}\phi^T\pa_{\mu}\phi-i\psibar\sigma.\pa\psi
-(\psi_a^TC(y^i)_{ab}\psi_b\phi^i+\hbox{h.c})-\tfrac{1}{4!}\lambda^{ijkl}\phi^i\phi^j\phi^k\phi^l,
\ee
where the Yukawa coupling $y^i$ is a $n_{\psi}\times n_{\psi}$ matrix, and $C$ is the charge conjugation matrix. As mentioned earlier, the features of interest first appear at three-loop order.
In this section we shall consider at three loops only the mixed Yukawa-scalar contributions. This is because the only relevant two-loop diagram in this case has only a simple pole, which simplifies the calculations. However we shall see later that the full calculation in fact has a very similar overall structure.  We start by describing the renormalisation process up to two loops. This allows us to define and discuss in some detail the renormalisation procedure for a general non-minimal scheme, allowing for the possibility of a hybrid scheme. We then turn to the three-loop calculation in the non-minimal scheme. Our explicit calculation is motivated by the (at least to us) surprising results for $\beta$-function contributions corresponding to 1PR contributions to the 2-point function, as obtained by coupling redefinition starting from $\MSbar$. It is however more natural pedagogically to start with the explicit 3-loop $\beta$-function computation of these terms (which we shall describe for conciseness as ``1PR2 terms'').  The method is quite complicated and somewhat indirect, so we begin by giving an overview of the procedure. Firstly we shall need the 2-point renormalisation constants corresponding to 3-loop 1PR contributions. However, the standard computation of the 2-point function only yields the symmetrised result. In order to obtain the corresponding antisymmetric part of the result and thus the separate contributions, we rely on an indirect calculation working back from the antisymmetric part of the anomalous dimension, which is obtained by coupling {\it and field}  redefinition from $\MSbar$. Next (in order to discuss the ``hybrid'' scheme where two and three point functions are treated differently) we consider the 3-loop 3-point diagrams corresponding to 1PR2 terms. The information we now have is sufficient to compute the required renormalisation constants. From these, we can compute the $\beta$-function coefficients; in a non-minimal scheme these are not simply derived from the simple poles in the renormalisation constants. We find a non-zero result for the ``1PR2'' contribution. Finally, we consider the predictions for these three-loop $\beta$-function terms, this time derived by coupling redefinition from $\MSbar$, and show that the prediction agrees exactly with the explicit calculation.

We start by regularising the theory by  replacing $\phi$, $\psi$ and $y$ by corresponding bare quantities $\phi_B$, $\psi_B$  and $y_B$, defined by
\begin{align}
\phi_B^i&=(\tZ_{\phi})^{ij}\phi^j,\nn
y^i_B&=(Z_{\phi}^{-1})^{ij}Z_{\psi}^{-1}y^jZ^{-1}_{\psi}+\yhat^i_B,\nn
(\lambda_B)^{ijkl}&=(Z_{\phi}^{-1})^{m(i}\lambda^{jkl)m},
\label{ybare}
\end{align}
with a similar expression for $\psi_B$. (However, it turns out that up to three loops, with the mixed scalar-Yukawa contributions to which we have restricted ourselves, we do not need to distinguish $\tZ_{\psi}$ from $Z_{\psi}$.)  
$\hat y^i_B$ denotes the terms corresponding to 1PI diagrams. This is all very standard except that we allow for the renormalisation constant $\tZ_{\phi}$ for $\phi$ to differ from that appearing in $y_B$, denoted $Z_{\phi}$. In $\MSbar$, for instance, they would be the same; but in a non-minimal scheme this is not mandatory--though $Z_{\phi}$ and $\tZ_{\phi}$ will differ only in their finite parts, or for 1PR contributions. Consequently we shall need to start by discussing two-point diagrams to fix $\tZ_{\phi}$ and then turn to three-point diagrams to determine $Z_{\phi}$ and $\hat y$; and, contrary to the usual $\MSbar$ calculation, we shall need to include 1PR three-point diagrams to compute $\tZ_{\phi}$ and $Z_{\phi}$.

We start by discussing the calculation up to two loops, which will give us the opportunity to introduce the various renormalisation schemes. The two-loop results may all be found in Ref.~\cite{JO1}. At one loop the relevant two-point and three-point diagrams are shown in Table \ref{1loop4d}.  They give

\begin{table}[h]
	\setlength{\extrarowheight}{1cm}
	\setlength{\tabcolsep}{24pt}
	\hspace*{-6.75cm}
	\centering
	\resizebox{6.7cm}{!}{
		\begin{tabular*}{20cm}{cccc}
			\begin{picture}(240,130) (161,-207)
			\SetWidth{0.5}
			\SetColor{Black}
			\SetWidth{1.0}
			\Arc(288,-142)(40.162,109,469)
			\Line[dash,dashsize=10](198,-142)(248,-142)
			\Line[dash,dashsize=10](328,-142)(378,-142)
			\SetColor{White}
			\Line(198,-78)(198,-206)
			\end{picture}
			&
			\begin{picture}(218,130) (161,-207)
			\SetWidth{0.5}
			\SetColor{Black}
			\Text(252,-106)[]{\Huge{\Black{$\tilde{Z}^{-1}_{\phi}$}}}
			\Text(334,-106)[]{\Huge{\Black{$(\tilde{Z}^{-1}_{\phi})^{T}$}}}
			\SetWidth{1.0}
			\Line[dash,dashsize=9.6](198,-142)(378,-142)
			\SetColor{White}
			\Line(198,-78)(198,-206)
			\SetColor{Black}
			\Line(272.002,-157.998)(303.998,-126.002)\Line(272.002,-126.002)(303.998,-157.998)
			\end{picture}
			&
			\begin{picture}(240,130) (161,-207)
			\SetWidth{0.5}
			\SetColor{Black}
			\SetWidth{1.0}
			\Arc(288,-142)(40.162,109,469)
			\Line[dash,dashsize=10](198,-142)(248,-142)
			\Line[dash,dashsize=10](328,-142)(378,-142)
			\Line(198,-78)(198,-206)
			\end{picture}
			&
			\begin{picture}(218,130) (161,-207)
			\SetWidth{0.5}
			\SetColor{Black}
			\Text(300,-106)[]{\Huge{\Black{$(\tilde{Z}^{-1}_{\phi})^{T}$}}}
			\SetWidth{1.0}
			\Line[dash,dashsize=9.6](198,-142)(378,-142)
			\Line(198,-78)(198,-206)
			\Vertex(198,-142){10}
			\Line(272.002,-157.998)(303.998,-126.002)\Line(272.002,-126.002)(303.998,-157.998)
			\end{picture}
			\\
			\hspace*{1cm}{\Huge $\tilde{G}^{(1)}_{1}$}
			&
			\hspace*{2cm}{\Huge $\tilde{G}^{(1)}_{2}$}
			&
			\hspace*{1cm}{\Huge $G^{(1)}_{1}$}
			&
			\hspace*{2cm}{\Huge $G^{(1)}_{2}$}
		\end{tabular*}
	}
	\caption{One-loop renormalization quantities for Scalar-Fermion theory}
	\label{1loop4d}	
\end{table}

\begin{align}
\tG^{(1)}_1&=g_1T_1,\nn
\tG^{(1)}_2&=2\tZ_1,\nn
G^{(1)}_1&=g_1T_{1y},\nn
G^{(1)}_2&=\tZ_{1y}+Z_{1y}
\label{Gdef}
\end{align}
where
\be
g_1=\frac{\gamma_{11}}{\epsilon}+\gamma_{10}+\ldots
\label{gdef}
\ee
(ignoring here $O(\epsilon)$ terms), and
\be
(T_1)^{ij}=\tr(y^iy^j).
\ee
We also introduce the notational conventions that for any 2-index quantity such as $T_1$, we denote $T_{1y}^i= y^j(T_1)_j{}^i$, and that $\tG^{(1)}_1$ etc represent two-point graphs, while $G^{(1)}_1$ etc represent three-point graphs.
$\tG^{(2)}_2$ and $G^{(2)}_2$ correspond to counterterm diagrams where we write $Z_{\phi}$, $\tZ_{\phi}$ in the form
\begin{align}
Z_{\phi}^{-1}&=1+Z_1+Z_2+Z_3+\ldots,\nn
\tZ_{\phi}^{-1}&=1+\tZ_1+\tZ_2+\tZ_3+\ldots,
\label{zphi}
\end{align}
where $Z_L$, $\tZ_L$ denote $L$-loop contributions. In this and other similar diagrams later, a ``blob'' at a vertex indicates a counterterm contribution from the corresponding bare coupling. Note that up to two loops $Z_1=Z_1^T$ and $Z_2=Z_2^T$ and similarly for $\tZ_1$, $\tZ_2$.
Explicit computation of the diagrams gives
\be
\gamma_{11}=\gamma_{10}=-2.
\label{gammas}
\ee
(here and elsewhere in the current section we suppress a standard factor of $(16\pi^2)^{-1}$ for each loop order in four dimensions).  However, the structure of the computation will be clearer if we refrain from inserting these explicit values. 
We then require that
\be
\tG^{(1)}_1+\tG^{(1)}_2=\hbox{finite}, \quad G^{(1)}_1+G^{(1)}_2=\hbox{finite}
\label{Gfin}
\ee
We can take in Eq.~\eqref{zphi}
\be
Z_1=z_1T_1,\quad \tZ_1=\tz_1T_1.
\label{Zdef}
\ee
We write
\be
z_1=\frac{\zeta_{11}}{\epsilon}+\zeta_{10},\quad
\tz_1=\frac{\tzeta_{11}}{\epsilon}+\tzeta_{10},
\label{ztilb}
\ee
and we readily see that Eq.~\eqref{Gfin} is satisfied if we take
\be
\tzeta_{11}=\zeta_{11}=-\tfrac12\gamma_{11}.
\label{zetasa}
\ee
Clearly the order of proceeding in principle  is first to determine $\tzeta_{11}$ from the two-point diagrams (which do not involve $\zeta_{11}$ at this order), and then $\zeta_{11}$ from the three-point diagrams; a rather trivial process here, but less so at higher orders. The finite parts in Eq.~\eqref{ztilb} are arbitrary, and therefore we can choose them to be different; for convenience we write
\be
\tzeta_{10}=\zeta_{10}+\delta_1.
\label{Zdefa}
\ee
We note here that the MOM scheme corresponds to taking $\zeta_{10}=-\tfrac12\gamma_{10}$ and the standard implementation of MOM also requires $\tzeta_{10}=\zeta_{10}$ and therefore $\delta_1=0$. 
\begin{table}[t]
	\setlength{\extrarowheight}{1cm}
	\setlength{\tabcolsep}{24pt}
	\hspace*{-8.5cm}
	\centering
	\resizebox{6.7cm}{!}{
		\begin{tabular*}{20cm}{cccc}
			\begin{picture}(240,130) (161,-207)
			\SetWidth{1.0}
			\SetColor{Black}
			\Arc[dash,dashsize=10](288,-142)(40.311,157,517)
			\Line[dash,dashsize=9.6](198,-142)(378,-142)
			\SetColor{White}
			\Line(198,-78)(198,-206)
			\end{picture}
			&
			\begin{picture}(307,130) (207,-207)
			\SetWidth{1.0}
			\SetColor{Black}
			\Arc(410,-142)(40,127,487)
			\Line[dash,dashsize=9](450,-142)(493,-142)
			\Arc(288,-142)(40,127,487)
			\Line[dash,dashsize=9](208,-142)(248,-142)
			\Line[dash,dashsize=9](328,-142)(368,-142)
			\SetColor{White}
			\Line(208,-78)(208,-206)
			\end{picture}
			&
			\begin{picture}(242,130) (175,-207)
			\SetWidth{0.5}
			\SetColor{Black}
			\Text(387,-107)[]{\Huge{\Black{$(\tilde{Z}^{-1}_{\phi})^{T}$}}}
			\Text(235,-173)[]{\Huge{\Black{$y_{B}$}}}
			\Text(341,-173)[]{\Huge{\Black{$y_{B}$}}}
			\SetWidth{1.0}
			\Arc(288,-142)(40.162,109,469)
			\Line[dash,dashsize=10](176,-142)(248,-142)
			\Line[dash,dashsize=10](328,-142)(416,-142)
			\Line(368.002,-157.998)(399.998,-126.002)\Line(368.002,-126.002)(399.998,-157.998)
			\Line(190.002,-157.998)(221.998,-126.002)\Line(190.002,-126.002)(221.998,-157.998)
			\Text(206,-107)[]{\Huge{\Black{$\tilde{Z}^{-1}_{\phi}$}}}
			\SetColor{White}
			\Line(176,-78)(176,-206)
			\end{picture}
			&
			\begin{picture}(218,130) (161,-207)
			\SetWidth{0.5}
			\SetColor{Black}
			\Text(252,-106)[]{\Huge{\Black{$\tilde{Z}^{-1}_{\phi}$}}}
			\Text(334,-106)[]{\Huge{\Black{$(\tilde{Z}^{-1}_{\phi})^{T}$}}}
			\SetWidth{1.0}
			\Line[dash,dashsize=9.6](198,-142)(378,-142)
			\SetColor{White}
			\Line(198,-78)(198,-206)
			\SetColor{Black}
			\Line(272.002,-157.998)(303.998,-126.002)\Line(272.002,-126.002)(303.998,-157.998)
			\end{picture}
			\\
			\hspace*{1cm}{\Huge $\tilde{G}^{(2)}_{1}$}
			&
			{\Huge $\tilde{G}^{(2)}_{2}$}
			&
			{\Huge $\tilde{G}^{(2)}_{3}$}
			&
			\hspace*{2cm}{\Huge $\tilde{G}^{(2)}_{4}$}
			\\
			&
			&
			&
			\\
			\begin{picture}(240,130) (161,-207)
			\SetWidth{1.0}
			\SetColor{Black}
			\Arc[dash,dashsize=10](288,-142)(40.311,157,517)
			\Line[dash,dashsize=9.6](198,-142)(378,-142)
			\Line(198,-78)(198,-206)
			\end{picture}
			&
			\begin{picture}(307,130) (207,-207)
			\SetWidth{1.0}
			\SetColor{Black}
			\Arc(410,-142)(40,127,487)
			\Line[dash,dashsize=9](450,-142)(493,-142)
			\Arc(288,-142)(40,127,487)
			\Line[dash,dashsize=9](208,-142)(248,-142)
			\Line[dash,dashsize=9](328,-142)(368,-142)
			\Line(208,-78)(208,-206)
			\end{picture}	
			&
			\begin{picture}(256,130) (161,-207)
			\SetWidth{0.5}
			\SetColor{Black}
			\Vertex(198,-142){10}
			\Vertex(248,-142){10}
			\Vertex(328,-142){10}
			\Text(390,-105)[]{\Huge{\Black{$(\tilde{Z}^{-1}_{\phi})^{T}$}}}
			\SetWidth{1.0}
			\Arc(288,-142)(40.162,109,469)
			\Line[dash,dashsize=10](198,-142)(248,-142)
			\Line[dash,dashsize=10](328,-142)(416,-142)
			\Line(198,-78)(198,-206)
			\Line(368.002,-157.998)(399.998,-126.002)\Line(368.002,-126.002)(399.998,-157.998)
			\end{picture}	
			&
			\begin{picture}(218,130) (161,-207)
			\SetWidth{0.5}
			\SetColor{Black}
			\Vertex(198,-142){10}
			\Text(294,-106)[]{\Huge{\Black{$(\tilde{Z}^{-1}_{\phi})^{T}$}}}
			\SetWidth{1.0}
			\Line[dash,dashsize=9.6](198,-142)(378,-142)
			\Line(198,-78)(198,-206)
			\Line(272.002,-157.998)(303.998,-126.002)\Line(272.002,-126.002)(303.998,-157.998)
			\end{picture}	
			\\
			\hspace*{1cm}{\Huge $G^{(2)}_{1}$}
			&
			{\Huge $G^{(2)}_{2}$}
			&
			{\Huge $G^{(2)}_{3}$}
			&
			\hspace*{2cm}{\Huge $G^{(2)}_{4}$}	
		\end{tabular*}
	}
	\caption{Two-loop renormalization quantities for Scalar-Fermion theory}
	\label{2loop4d}	
\end{table}

At two loops the relevant two-point and three-point diagrams are shown in Table \ref{2loop4d}. They give
contributions denoted $\tG_{\alpha}^{(3)}$, given by
\begin{align}
\tG^{(2)}_1&=g_2T_2,\nn
\tG^{(2)}_2&=g_1^2T_1^2,\nn
\tG^{(2)}_3&=2(\tz_1+z_1)g_1T_1^2,\nn
\tG^{(2)}_4&=2\tZ_2,\nn
G^{(2)}_1&=g_2T_{2y},\nn
G^{(2)}_2&=g_1^2(T_1^2)_y,\nn
G^{(2)}_3&=(\tz_1+3z_1)g_1(T_1^2)_y,\nn
G^{(2)}_4&=\tZ_{2y}+Z_{2y}
\label{Gdef}
\end{align}
where $g_1$ is defined in Eq.~\eqref{gdef}, $g_2$ is defined by
\be
g_2=\frac{\gamma_{21}}{\epsilon}+\gamma_{20}+\ldots
\label{gdeftwo}
\ee
(once again ignoring $O(\epsilon)$ terms), and
\be
(T_2)^{ij}=\lambda^{iklm}\lambda^{klmj}.
\ee
As mentioned earlier, we note that the 1PI two-loop diagram $\tG^{(2)}_1$ has only a single pole. Explicit computation of the diagrams gives
\be
\gamma_{21}=-\tfrac{1}{12},\quad \gamma_{20}=-\tfrac{13}{48}.
\label{gammas}
\ee
We note here the important fact that contributions from $\tZ_{\phi}$ only appear on the external legs of the diagrams.
The reason for this is that on any internal line of any diagram, the contributions from $\tZ_{\phi}$ at the vertices cancel those from $\tZ_{\phi}^{-1}$ arising from the propagators. We shall discuss the significance of this later.
We now require that
\be
\tG^{(2)}_1+\tG^{(2)}_2+\tG^{(2)}_3+\tG^{(2)}_4=\hbox{finite}, \quad G^{(2)}_1+G^{(2)}_2+G^{(2)}_3+G^{(2)}_4=\hbox{finite.}
\label{Gfina}
\ee
We can take in Eq.~\eqref{zphi}
\be
Z_2=z_2T_2+x_2T_1^2,
\label{Zdefb}
\ee
with a similar expression for  $\tZ_2$. We write
\be
z_2=\frac{\zeta_{21}}{\epsilon}+\zeta_{20},\quad
\tz_2=\frac{\tzeta_{21}}{\epsilon}+\tzeta_{20},
\label{ztild}
\ee
where again we write
\be
 \tzeta_{20}=\zeta_{20}+\delta_2.
\label{Zdefd}
\ee
Once again we determine the pole terms in $\tZ_2$ and $Z_2$ by imposing finiteness in Eq.~\eqref{Gfina} first upon the two-point and then upon the three-point graphs. We find that finiteness is assured by taking
\be
 \zeta_{21}=\tzeta_{21}=-\tfrac12\gamma_{21},
\label{zetasc}
\ee
together with
\be
x_2=\tfrac32(z_1^2-\zeta_{10}^2)+\xi_4,\quad \tx_2=\tfrac32(z_1^2-\zeta_{10}^2)+(z_1-\zeta_{10})\delta_1+\txi_4,
\label{xtwo}
\ee
and $z_1$ as in Eq.~\eqref{ztilb}. The finite quantities $\zeta_{20}$, $\tzeta_{20}$, $\xi_4$ and $\txi_4$ are arbitrary.
The MOM scheme corresponds to taking 
\be
\zeta_{20}=-\tfrac12\gamma_{20} 
\ee
in Eq.~\eqref{ztild}; and the standard implementation would also have $\delta_2=0$ in Eq.~\eqref{Zdefd}.
We shall henceforth adopt the convention of omitting the tilde
on quantities in $\tZ_{\phi}$ if they are the same as the corresponding quantities in $Z_{\phi}$. An important consequence  of the afore-mentioned cancellation of $\tZ_{\phi}$ on internal lines is that the pole parts of counterterms for 1PI 3-point and 2-point terms depend only on the poles in the diagrams, together with the finite parts in $Z_{\phi}$. In other words they depend only on $\gamma_{11}$, etc, and $\zeta_{10}$, etc, and not on $\tzeta_{10}$, etc (or, equivalently, on $\delta_1$, etc). This means that the scheme dependence of the 1PI contributions to the $\beta$-function is determined purely by the choice of finite parts in $Z_{\phi}$; the finite parts of $\tZ_{\phi}$ play no role in this. Another consequence is that all the pole terms for 1PI contributions to $\tZ_{\phi}$ are exactly as in $Z_{\phi}$; they differ only in the finite contribution.
  In fact in the present case, at two loops the pole parts of $Z_2$ and $\tZ_2$ only depend on the diagram results $\gamma_{21}$, and not on $\zeta_{10}$; but this is a consequence of the absence of a double pole in $g_2$ in Eq.~\eqref{gdeftwo}. We shall see the more general situation in the six-dimensional theory discussed later. However, we do see in Eq.~\eqref{xtwo} that the 1PR counterterm $\tx_2$ has acquired a $\delta_1$-dependent pole.

We may now use the results from the Appendix to compute the one and two loop $\beta$-functions. As explained in the Appendix, in a non-minimal scheme the $\beta$-function has $\epsilon$-dependent terms and we may write
\be
\beta=-\tfrac12y\epsilon+(b_1+b_{01}\epsilon)
T_{1y}+(b_2+b_{02}\epsilon)T_{2y}+\ldots
\ee
where we find from Eqs.~\eqref{betares}, \eqref{Zres} that 
\be
b_1=\zeta_{11},\quad b_{01}=\zeta_{01},\quad b_2=2\zeta_{21},\quad b_{02}=2\zeta_{20};
\label{betrestwo}
\ee
the ellipsis indicates that we are neglecting terms involving loops on fermion lines,
together with 1PI contributions to $g_{21}$ (in the notation of Eq.~\eqref{gb}). For future convenience, with a slight abuse of terminology we shall use the phrase ``$\beta$-function'' to refer to the $\epsilon$-independent portion. We note that as in the usual $\MSbar$ case, the $\beta$-function is derived from the simple poles in the 1PI two and three point counterterms (though at higher loop order, and in Section 3 even at two loops, lower-order finite terms will also appear). It is however important to note that despite the simple pole in the 1PR term $x_2$ in Eq.~\eqref{xtwo}, there is no corresponding contribution to the $\beta$-function. We see that in this case and up to this order the $\beta$-function is scheme-independent, having no dependence on our choice of $\zeta_{10}$. This is consistent with the deduction from coupling redefinitions. From Eqs.~\eqref{Zres}, \eqref{redexp},  we expect that in general the change of scheme from $\MSbar$ to a general non-minimal scheme may be effected up to two loops by
\be
\delta y=-\zeta_{10}T_{1y}.
\label{yredef}
\ee%
It is easy to check using Eqs.~\eqref{betrestwo}, \eqref{redeflow},  that at two loops
\be
\delta\beta^i=0,
\ee
in agreement with the scheme independence of $b_2$ in Eq.~\eqref{betrestwo}.
In fact there are also 
cross-terms involving products of pairs of $T_1$, $T_{\psi}=y^iy^i$ in $g_{21}$ but again the corresponding potential contributions to the $\beta$-function cancel out.
 
We now turn to the explicit three-loop calculation of 1PR2 contributions within the non-minimal scheme. The 1PR2 $\beta$-function coefficients are determined by 3-loop simple poles in $Z_{\phi}$ (together with lower-loop terms, as we shall see). As we commented earlier, the simple poles for 1PI contributions to $Z_{\phi}$,  $\tZ_{\phi}$ are identical. However, this is not the case for the 1PR2 simple pole contributions beyond two loops, since they are affected by the lower-order finite differences; therefore we need to consider the derivations of 1PR contributions to $Z_{\phi}$ and $\tZ_{\phi}$ separately in order to derive the $\beta$-function. We must start with $\tZ_{\phi}$, since this result will feed into the calculation for $Z_{\phi}$. We therefore consider the three-loop 1PR2 2-point diagrams, which are shown in Table \ref{3loop4d2} and give

\begin{table}[t]
	\setlength{\extrarowheight}{1cm}
	\setlength{\tabcolsep}{24pt}
	\hspace*{-11cm}
	\centering
	\resizebox{6cm}{!}{
		\begin{tabular*}{20cm}{ccccc}
			\begin{picture}(304,130) (207,-207)
			\SetWidth{1.0}
			\SetColor{Black}
			\Arc[dash,dashsize=10](410,-142)(40.311,157,517)
			\Line[dash,dashsize=9.6](208,-142)(248,-142)
			\Arc(288,-142)(40,127,487)
			\Line[dash,dashsize=9.6](328,-142)(513,-142)
			\SetColor{White}
			\Line(208,-78)(208,-206)
			\end{picture}
			&
			\begin{picture}(307,130) (207,-207)
			\SetWidth{1.0}
			\SetColor{Black}
			\Arc[dash,dashsize=10](288,-142)(40.311,157,517)
			\Line[dash,dashsize=9.6](208,-142)(369,-142)
			\Arc(410,-142)(40,127,487)
			\Line[dash,dashsize=9.6](450,-142)(513,-142)
			\SetColor{White}
			\Line(208,-78)(208,-206)
			\end{picture}
			&
			\begin{picture}(242,130) (175,-207)
			\SetWidth{0.5}
			\SetColor{Black}
			\Text(397,-107)[]{\Huge{\Black{$(\tilde{Z}^{-1}_{\phi})^{T}$}}}
			\Vertex(248,-142){10}
			\Vertex(328,-142){10}
			\SetWidth{1.0}
			\Arc(288,-142)(40.162,109,469)
			\Line[dash,dashsize=10](176,-142)(248,-142)
			\Line[dash,dashsize=10](328,-142)(416,-142)
			\Line(368.002,-157.998)(399.998,-126.002)\Line(368.002,-126.002)(399.998,-157.998)
			\Line(190.002,-157.998)(221.998,-126.002)\Line(190.002,-126.002)(221.998,-157.998)
			\Text(206,-107)[]{\Huge{\Black{$\tilde{Z}^{-1}_{\phi}$}}}
			\SetColor{White}
			\Line(176,-78)(176,-206)
			\end{picture}
			&
			\begin{picture}(240,130) (161,-207)
			\SetWidth{1.0}
			\SetColor{Black}
			\Vertex(248,-142){10}
			\Vertex(328,-142){10}
			\Arc[dash,dashsize=10](288,-142)(40.311,157,517)
			\Line[dash,dashsize=9.6](188,-142)(400,-142)
			\Line(352.002,-157.998)(383.998,-126.002)\Line(352.002,-126.002)(383.998,-157.998)
			\Line(190.002,-157.998)(221.998,-126.002)\Line(190.002,-126.002)(221.998,-157.998)
			\Text(206,-107)[]{\Huge{\Black{$\tilde{Z}^{-1}_{\phi}$}}}
			\Text(370,-107)[]{\Huge{\Black{$(\tilde{Z}^{-1}_{\phi})^{T}$}}}
			\SetColor{White}
			\Line(188,-78)(198,-206)
			\end{picture}
			&
			\begin{picture}(218,130) (161,-207)
			\SetWidth{0.5}
			\SetColor{Black}
			\Text(259,-106)[]{\Huge{\Black{$\tilde{Z}^{-1}_{\phi}$}}}
			\Text(329,-106)[]{\Huge{\Black{$(\tilde{Z}^{-1}_{\phi})^{T}$}}}
			\SetWidth{1.0}
			\Line[dash,dashsize=9.6](198,-142)(378,-142)
			\Line(272.002,-157.998)(303.998,-126.002)\Line(272.002,-126.002)(303.998,-157.998)
			\SetColor{White}
			\Line(198,-78)(198,-206)
			\end{picture}
			\\
			{\Huge $\tilde{G}^{(3)}_{1}$}
			&
			{\Huge $\tilde{G}^{(3)}_{2}$}
			&
			{\Huge $\tilde{G}^{(3)}_{3}$}
			&
			\hspace*{1cm}{\Huge $\tilde{G}^{(3)}_{4}$}
			&
			\hspace*{1.5cm}{\Huge $\tilde{G}^{(3)}_{5}$}
		\end{tabular*}
	}
	\caption{Three-loop, two-point renormalization quantities}
	\label{3loop4d2}	
\end{table}

\begin{align}
\tG^{(3)}_{1}&=g_1g_2T_1T_2,\nn
\tG^{(3)}_{2}&=g_1g_2T_2T_1,\nn
\tG^{(3)}_3&=(Z_2+\tZ_2)g_1T_1+g_1T_1(Z_2+\tZ_2)\nn
\tG^{(3)}_4&=(Z_1+\tZ_1)g_2T_2+g_2T_2(Z_1+\tZ_1)\nn
\tG^{(3)}_5&=\tZ_1\tZ_2+\tZ_2\tZ_1+\tZ_3+\tZ_3^T,
\label{Ddiag}
\end{align}
where $g_1$, $g_2$ are as defined in Eqs.~\eqref{gdef},
\eqref{gdeftwo}, $Z_{\phi}$, $\tZ_{\phi}$ are expanded as in Eq.~\eqref{zphi}, and we write
\be
\tZ_3=\tz_3T_1T_2+\tz_3'T_2T_1+\ldots
\label{tzexp}
\ee
where
\be
\tz_3=\frac{\tzeta_{32}}{\epsilon^2}+\frac{\tzeta_{31}}{\epsilon}
+\tzeta_{30},\quad\tz'_3=\frac{\tzeta'_{32}}{\epsilon^2}+\frac{\tzeta'_{31}}{\epsilon}
+\tzeta'_{30},\quad 
\label{zthree}
\ee
and
\be
\tzeta_{30}=\zeta_{30}+\delta_{30},\quad
\tzeta'_{30}=\zeta'_{30}+\delta'_{30}.
\ee
Here $\zeta_{30}$, $\zeta'_{30}$ are the corresponding quantities in $Z_3$, expanded in a similar manner to Eqs.~\eqref{tzexp}, \eqref{zthree}.
Of course there are also 1PI three-loop two-point diagrams, but as we have explained, we are not concerned with these here and do not consider them further. Since $\tZ_3$ appears in Eq.~\eqref{Ddiag} in symmetrised form, only the sum $\tzeta_{31}+\tzeta_{31}'$ will be defined. It might seem natural to assume $\tzeta_{31}=\tzeta_{31}'$, but in fact this is not the case, as we shall see.
We have
\be
\sum \tG_{\alpha}^{(3)}=\hbox{finite}.
\ee
We find, inserting Eqs.~\eqref{gdef}, \eqref{Zdef}-\eqref{Zdefa}, \eqref{gdeftwo},  \eqref{Zdefb} and \eqref{Zdefd}
into Eq.~\eqref{Ddiag},
\begin{align}
\tzeta_{32}&=\tzeta_{32}'=\tfrac32\zeta_{11}\zeta_{21},\nn
\tzeta_{31}+\tzeta_{31}'&=3C+\zeta_{11}\delta_2
+\zeta_{21}\delta_1,
\label{zzprimex}
\end{align}
where
\be
C=\zeta_{11}\zeta_{20}+\zeta_{21}\zeta_{10}.
\ee
The appearance of $C$ in the three-loop simple pole is of course natural, since it is just the simple pole in $\tG^{(3)}_{1}$ or  $\tG^{(3)}_{2}$ in Eq.~\eqref{Ddiag}. We parametrise $\tzeta_{31}$, $\tzeta_{31}'$ as
\begin{align}
\tzeta_{31}&=\tfrac32C+\tfrac12(\zeta_{11}\delta_2
+\zeta_{21}\delta_1)+\xi,\nn
\tzeta'_{31}&=\tfrac32C+\tfrac12(\zeta_{11}\delta_2
+\zeta_{21}\delta_1)-\xi.
\label{zzprime}
\end{align}
We now need to compute $\xi$. For this we use the indirect method outlined in Appendix B, which will enable us to reconstruct $\xi$ from the asymmetric part of the anomalous dimension as obtained by coupling and field redefinition.  This enables us to deduce the individual simple pole coefficients $\tzeta_{31}$, $\tzeta_{31}'$. We write
\be
\gamma_{\phi}=\gamma_1T_1+\gamma_2T_2+\gamma_{3}T_1T_2+\gamma'_{3}T_2T_1+\ldots
\ee
where the ellipsis indicates irrelevant three-loop terms as well as higher-order contributions. 
Inserting Eq.~\eqref{tzexp} with the results of Eq.~\eqref{Zres} into Eq.~\eqref{gamthree}, we find
\be
\gamma_{\phi}^{(3)}=\epsilon(3\tz_3-\tz_1\tz_2)T_1T_2+\epsilon(3\tz'_3-2\tz_1\tz_2)T_2T_1
\ee
from which we deduce
\begin{align}
\gamma_{3}&=3\tzeta_{31}-(\zeta_{21}\tzeta_{10}+\tzeta_{20}\zeta_{11}),\nn
\gamma'_{3}&=3\tzeta'_{31}-2(\zeta_{21}\tzeta_{10}+\tzeta_{20}\zeta_{11}),
\label{gammone}
\end{align}
and also verify the double poles in Eq.~\eqref{zzprimex}. The difference $\gamma_3-\gamma'_3$ may in general also be computed by making the appropriate coupling and field redefinition from $\MSbar$. In fact all we need in this case  is the field redefinition $\phi'=\Omega\phi$  where as we see from Eq.~\eqref{redexp}
\be
\delta \Omega =\tzeta_{10}T_1+\tzeta_{20}T_2.
\ee
Then from the symmetry of $\gamma_{\phi}$ in $\MSbar$, we deduce from Eq.~\eqref{gamredef}  in Appendix B 
\be
\gamma_3-\gamma_3'=2(\gamma^{\MSbar}_2\tzeta_{10}-\gamma^{\MSbar}_1\tzeta_{20})
=2(2\zeta_{21}\tzeta_{10}-\zeta_{11}\tzeta_{20}).
\label{gammtwo}
\ee
Although we emphasise here that the anomalous dimension results in Eq.~\eqref{gammtwo} are evaluated in $\MSbar$, the one-loop term is of course automatically scheme-independent, and this particular two-loop term also happens to be scheme-independent too. Then  combining Eqs.~\eqref{gammone}, \eqref{gammtwo}, we deduce
\be
\xi=\tfrac12(\tzeta_{31}-\tzeta'_{31})=\tfrac12(\zeta_{21}\tzeta_{10}-\zeta_{11}\tzeta_{20}).
\label{xires}
\ee
On insertion into Eq.~\eqref{zzprime}, this gives
\begin{align}
\tzeta_{31}&=C+\zeta_{21}\tzeta_{10},\nn
\tzeta'_{31}&=C+\zeta_{11}\tzeta_{20}.
\label{zzprimen}
\end{align}
It is clear from the details of the calculation that this sort of asymmetry will only affect 1PR contributions to the $\gamma$-function.
We can now use these results in the computation of the 1PR 3-loop contributions to $Z_{\phi}^{-1}$ in Eq.~\eqref{zphi} (which will ultimately determine the $\beta$-function via Eq.~\eqref{ybare}). For this we need to consider the 1PR three-point diagrams depicted in Table \ref{3loop4d3} corresponding to the 1PR two-point diagrams of Table \ref{3loop4d2}. These diagrams give contributions to the 3-point function denoted $G_{\alpha}^{(3)}$, given by

\begin{table}[t]
	\setlength{\extrarowheight}{1cm}
	\setlength{\tabcolsep}{24pt}
	\hspace*{-10.5cm}
	\centering
	\resizebox{6cm}{!}{
		\begin{tabular*}{20cm}{ccccc}
			\begin{picture}(304,130) (207,-207)
			\SetWidth{1.0}
			\SetColor{Black}
			\Arc[dash,dashsize=10](410,-142)(40.311,157,517)
			\Line[dash,dashsize=9.6](208,-142)(248,-142)
			\Arc(288,-142)(40,127,487)
			\Line[dash,dashsize=9.6](328,-142)(513,-142)
			\Line(208,-78)(208,-206)
			\end{picture}
			&
			\begin{picture}(307,130) (207,-207)
			\SetWidth{1.0}
			\SetColor{Black}
			\Arc[dash,dashsize=10](288,-142)(40.311,157,517)
			\Line[dash,dashsize=9.6](208,-142)(369,-142)
			\Arc(410,-142)(40,127,487)
			\Line[dash,dashsize=9.6](450,-142)(513,-142)
			\Line(208,-78)(208,-206)
			\end{picture}
			&
			\begin{picture}(256,130) (161,-207)
			\SetWidth{0.5}
			\SetColor{Black}
			\Text(390,-106)[]{\Huge{\Black{$(\tilde{Z}^{-1}_{\phi})^{T}$}}}
			\SetWidth{1.0}
			\Arc(288,-142)(40.162,109,469)
			\Vertex(198,-142){10}
			\Vertex(248,-142){10}
			\Vertex(328,-142){10}
			\Line[dash,dashsize=10](198,-142)(248,-142)
			\Line[dash,dashsize=10](328,-142)(416,-142)
			\Line(368.002,-157.998)(399.998,-126.002)\Line(368.002,-126.002)(399.998,-157.998)
			\Line(198,-78)(198,-206)
			\end{picture}
			&
			\begin{picture}(240,130) (161,-207)
			\SetWidth{1.0}
			\SetColor{Black}
			\Arc[dash,dashsize=10](288,-142)(40.311,157,517)
			\Line[dash,dashsize=9.6](198,-142)(400,-142)
			\Line(352.002,-157.998)(383.998,-126.002)\Line(352.002,-126.002)(383.998,-157.998)
			\Text(375,-105)[]{\Huge{\Black{$(\tilde{Z}^{-1}_{\phi})^{T}$}}}
			\Vertex(198,-142){10}
			\Vertex(248,-142){10}
			\Vertex(328,-142){10}
			\Line(198,-78)(198,-206)
			\end{picture}
			&
			\begin{picture}(218,130) (161,-207)
			\SetWidth{0.5}
			\SetColor{Black}
			\Text(290,-106)[]{\Huge{\Black{$(\tilde{Z}^{-1}_{\phi})^{T}$}}}
			\SetWidth{1.0}
			\Line[dash,dashsize=9.6](198,-142)(378,-142)
			\Vertex(198,-142){10}
			\Line(272.002,-157.998)(303.998,-126.002)\Line(272.002,-126.002)(303.998,-157.998)
			\Line(198,-78)(198,-206)
			\end{picture}
			\\
			{\Huge $G^{(3)}_{1}$}
			&
			{\Huge $G^{(3)}_{2}$}
			&
			{\Huge $G^{(3)}_{3}$}
			&
			\hspace*{1cm}{\Huge $G^{(3)}_{4}$}
			&
			\hspace*{1.5cm}{\Huge $G^{(3)}_{5}$}
		\end{tabular*}
	}
	\caption{Three-loop, three-point renormalization quantities}
	\label{3loop4d3}	
\end{table}

\begin{align}
G^{(3)}_{1}&=g_1g_2(T_1T_2)_y\nn
G^{(3)}_{2}&=g_1g_2(T_2T_1)_y\nn
G^{(3)}_3&=2g_1(Z_2T_1)_y+[g_1(T_1Z_2+T_1\tZ_2)]_y\nn
G^{(3)}_4&=2g_2(Z_1T_2)_y+g_2[T_2(Z_1+\tZ_1)]_y\nn
G^{(3)}_5&=(Z_1\tZ_2+Z_2\tZ_1+Z_3+\tZ_3^T)_y
\label{GT}
\end{align}
We have
\be
\sum G^{(3)}_{\alpha}=\hbox{finite}.
\ee
We write, analogously to Eqs.~\eqref{tzexp}, \eqref{zthree},
\be
Z_3=z_3T_1T_2+z_3'T_2T_1+\ldots
\ee
with
\be
z_3=\frac{\zeta_{32}}{\epsilon^2}+\frac{\zeta_{31}}{\epsilon}
+\zeta_{30},\quad
z_3'=\frac{\zeta'_{32}}{\epsilon^2}+\frac{\zeta'_{31}}{\epsilon}
+\zeta'_{30}.
\ee
Inserting Eqs.~\eqref{gdef}, \eqref{Zdef}-\eqref{Zdefa}, \eqref{gdeftwo},  \eqref{Zdefb}, \eqref{Zdefd}, \eqref{zzprime} into Eq.~\eqref{GT} , we obtain
\begin{align}
\zeta_{32}&=\zeta_{32}'=\tfrac32\zeta_{11}\zeta_{21},\nn
\zeta_{31}&=\tfrac32C+\tfrac12(\zeta_{11}\delta_2-\zeta_{21}\delta_1)+\xi,\nn
\zeta'_{31}&=\tfrac32C+\tfrac12(\zeta_{11}\delta_2-\zeta_{21}\delta_1)-\xi,
\end{align}
where $\delta_1$, $\delta_2$ are defined in Eqs.~\eqref{Zdefa}, \eqref{Zdefd}.
We now deduce using Eq.~\eqref{xires}
\begin{align}
\zeta_{31}&=C+\zeta_{21}\zeta_{10},\nn
\zeta'_{31}&=C+\zeta_{11}\zeta_{20}.
\label{zfin}
\end{align}
Note that there is no dependence on $\tzeta_{10}$, $\tzeta_{20}$ in $\zeta_{31}$, $\zeta_{31}'$; however, as expected, comparing Eqs.~\eqref{zzprimen}, \eqref{zfin} we see that $\tzeta_{31}=\zeta_{31}$ and $\zeta'_{31}=\zeta'_{31}$ in the case when $\tzeta_{10}=\zeta_{10}$, $\tzeta_{20}=\zeta_{20}$. 
Finally we are able to compute the corresponding $\beta$-function contributions. Using Eq.~\eqref{betrestwo}, \eqref{betares}, \eqref{Zres}, and writing $b_3$, $b_3'$ for the coefficients of $y^j(T_1T_2)_j{}^i$
and $y^j(T_2T_1)_j{}^i$, respectively, in the $\beta$-function, we find
\begin{align}
b_3&=3\zeta_{31}-4C,\nn
b'_3&=3\zeta'_{31}-5C,
\label{betfinb}
\end{align}
from which we obtain using Eq.~\eqref{zfin},
\be
b_3=-b'_3=b_{2}\zeta_{10}-b_1\zeta_{20},
\label{betfin}
\ee
where the lower-order $\beta$-function coefficients are as defined in Eq.~\eqref{betrestwo}.
We finally see that the 1PR2 $\beta$-function coefficients are non-zero. We shall now show that this agrees with the prediction obtained by a scheme change from $\MSbar$.
The change of renormalisation scheme corresponds to a coupling redefinition which can be derived from Eq.~\eqref{redexp} using Eqs.~\eqref{Zres}. We find
\be
\delta y^i=-\zeta_{10}(T_1)^{ij}y^j-\zeta_{20}(T_2)^{ij}y^j,
\label{yredef}
\ee%
and
\be
\delta \lambda^{ijkl}=-\zeta_{10}(T_1)^{m(i}\lambda^{jkl)m}.
\ee
We now use the general results for the lowest-order effect of a scheme change given in Eq.~\eqref{redeflow}. Using the results for the one and two-loop $\MSbar$ $\beta_y^i$ from Eqs.~\eqref{betrestwo},
Eq.~\eqref{redeflow} gives the three-loop change
\be
\delta b_3=-\delta b_3'=b_{2}\zeta_{10}-b_{1}\zeta_{20}
\label{deltwo}
\ee
using the notation introduced in Eqs.~\eqref{betfinb}, \eqref{betfin}. This indeed agrees with the result obtained by explicit calculation (of course  we used part of the scheme change prediction, for the anomalous dimension, in the explicit calculation, so to be more precise we should simply say that the explicit results and scheme change predictions are consistent). 
We shall return to a discussion of the remaining 1PR2 contributions after Sect. 3.

\section{Scalar theory in six dimensions}
In this section we shall consider $\phi^3$ theory in six dimensions. The Lagrangian is given by
\be
L=\tfrac12\pa_{\mu}\phi^i\pa_{\mu}\phi^i+\tfrac{1}{3!}g^{ijk}\phi^i\phi^j\phi^k,
\ee
where $i=1\ldots N$.
We shall follow as far as possible the calculation of the previous section, using similar notation, in order to emphasise the close similarities in structure. The basic RG results for this theory have been known for some time\cite{jag1,jag2}; but here we pick out the distinctive features of our approach. Following the previous section, we again distinguish the $\tZ_{\phi}$ which appears in $\phi_B=\tZ_{\phi}\phi$ from the $Z_{\phi}$ which appears in the bare coupling $g_B$, i.e.
\be
g^{ijk}_B=(Z^{-1}_{\phi})^{l(i}g^{jk)l}+\hat g^{ijk}_B,
\label{gbdef}
\ee
where $Z_{\phi}^{-1}$, $\tZ_{\phi}^{-1}$ are again expanded as in Eq.~\eqref{zphi}. $\hat g^{ijk}_B$ again denotes the terms corresponding to 1PI diagrams, together with ``cross-terms'' resulting from 1PR structures on two or more ``legs'' of $g^{ijk}$.
$\tZ_{\phi}$ is of course determined by the two-point diagrams. The calculation up to two loops is very similar to that described for the scalar-fermion theory in Section 2, so we shall present both the one-loop and two-loop results at the same time; but we emphasise that the calculational procedure follows Section 2; at one loop, $\tZ_1$ is  determined from 2-point graphs, followed by $Z_1$ from 3-point graphs; then similarly for $\tZ_2$ followed by $Z_2$. Up to two loops the relevant graphs are depicted in Table \ref{12loop6d}  and give

\begin{table}[t]
	\setlength{\extrarowheight}{1cm}
	\setlength{\tabcolsep}{24pt}
	\hspace*{-9cm}
	\centering
	\resizebox{6.7cm}{!}{
		\begin{tabular*}{20cm}{cccc}
			\begin{picture}(234,130) (241,-207)
			\SetWidth{1.0}
			\SetColor{Black}
			\Line(278,-142)(319,-142)
			\Line(399,-142)(439,-142)
			\Arc(359,-142)(39.598,135,495)
			\SetColor{White}
			\Line(278,-78)(278,-206)
			\end{picture}
			&
			\hspace*{-1cm}\begin{picture}(199,130) (241,-207)
			\SetWidth{1.0}
			\SetColor{Black}
			\Line(278,-142)(439,-142)
			\Text(338,-110)[]{\Huge{\Black{$\tilde{Z}^{-1}_{\phi}$}}}
			\Text(408,-110)[]{\Huge{\Black{$(\tilde{Z}^{-1}_{\phi})^{T}$}}}
			\Line(352.002,-157.998)(383.998,-126.002)\Line(352.002,-126.002)(383.998,-157.998)
			\SetColor{White}
			\Line(278,-78)(278,-206)
			\end{picture}
			&
			\begin{picture}(234,130) (241,-207)
			\SetWidth{1.0}
			\SetColor{Black}
			\Line(278,-142)(319,-142)
			\Line(399,-142)(439,-142)
			\Arc(359,-142)(39.598,135,495)
			\Line(278,-78)(278,-206)
			\end{picture}
			&
			\begin{picture}(199,130) (241,-207)
			\SetWidth{1.0}
			\SetColor{Black}
			\Line(278,-142)(439,-142)
			\Vertex(278,-142){10}
			\Text(378,-105)[]{\Huge{\Black{$(\tilde{Z}^{-1}_{\phi})^{T}$}}}
			\Line(278,-78)(278,-206)
			\Line(352.002,-157.998)(383.998,-126.002)\Line(352.002,-126.002)(383.998,-157.998)
			\end{picture}
			\\
			\hspace*{0.75cm}{\Huge $\tilde{G}^{(1)}_{1}$}
			&
			\hspace*{1cm}{\Huge $\tilde{G}^{(1)}_{2}$}
			&
			\hspace*{1cm}{\Huge $G^{(1)}_{1}$}
			&
			\hspace*{2.5cm}{\Huge $G^{(1)}_{2}$}
			\\
			&
			&
			&
			\\
			\begin{picture}(234,130) (241,-207)
			\SetWidth{1.0}
			\SetColor{Black}
			\Line(278,-142)(319,-142)
			\Line(399,-142)(439,-142)
			\Arc(359,-142)(39.598,135,495)
			\Arc(359,-92.857)(39.143,-140.034,-39.966)
			\SetColor{White}
			\Line(278,-78)(278,-206)
			\end{picture}
			&
			\begin{picture}(353,130) (172,-207)
			\SetWidth{1.0}
			\SetColor{Black}
			\Arc(288,-142)(39.598,135,495)
			\Arc(410,-142)(39.598,135,495)
			\Line(208,-142)(248,-142)
			\Line(328,-142)(370,-142)
			\Line(449,-142)(489,-142)
			\SetColor{White}
			\Line(208,-78)(208,-206)
			\end{picture}
			&
			\begin{picture}(211,130) (255,-207)
			\SetWidth{1.0}
			\SetColor{Black}
			\Vertex(319,-142){10}
			\Vertex(399,-142){10}
			\Line(256,-142)(319,-142)
			\Line(399,-142)(465,-142)
			\Arc(359,-142)(39.598,135,495)
			\Text(288,-105)[]{\Huge{\Black{$\tilde{Z}^{-1}_{\phi}$}}}
			\Text(437,-105)[]{\Huge{\Black{$(\tilde{Z}^{-1}_{\phi})^{T}$}}}
			\Line(416.002,-157.998)(447.998,-126.002)\Line(416.002,-126.002)(447.998,-157.998)
			\Line(272.002,-157.998)(303.998,-126.002)\Line(272.002,-126.002)(303.998,-157.998)
			\SetColor{White}
			\Line(256,-78)(256,-206)
			\end{picture}
			&
			\begin{picture}(199,130) (241,-207)
			\SetWidth{1.0}
			\SetColor{Black}
			\Line(278,-142)(439,-142)
			\Text(328,-110)[]{\Huge{\Black{$\tilde{Z}^{-1}_{\phi}$}}}
			\Text(418,-110)[]{\Huge{\Black{$(\tilde{Z}^{-1}_{\phi})^{T}$}}}
			\Line(352.002,-157.998)(383.998,-126.002)\Line(352.002,-126.002)(383.998,-157.998)
			\SetColor{White}
			\Line(278,-78)(278,-206)
			\end{picture}
			\\
			\hspace{0.5cm}{\Huge $\tilde{G}^{(2)}_{1}$}
			&
			\hspace{0.5cm}{\Huge $\tilde{G}^{(2)}_{2}$}
			&
			\hspace{0.5cm}{\Huge $\tilde{G}^{(2)}_{3}$}
			&
			\hspace{2.5cm}{\Huge $\tilde{G}^{(2)}_{4}$}
		\end{tabular*}
	}
\end{table}

\begin{table}[t]
	\setlength{\extrarowheight}{1cm}
	\setlength{\tabcolsep}{10pt}
	\hspace*{-10cm}
	\centering
	\resizebox{7cm}{!}{
		\begin{tabular*}{20cm}{ccccc}
			\begin{picture}(234,130) (241,-207)
			\SetWidth{1.0}
			\SetColor{Black}
			\Line(278,-142)(319,-142)
			\Line(399,-142)(439,-142)
			\Arc(359,-142)(39.598,135,495)
			\Arc(359,-92.857)(39.143,-140.034,-39.966)
			\Line(278,-78)(278,-206)
			\end{picture}
			&
			\begin{picture}(353,130) (172,-207)
			\SetWidth{1.0}
			\SetColor{Black}
			\Arc(288,-142)(39.598,135,495)
			\Arc(410,-142)(39.598,135,495)
			\Line(208,-142)(248,-142)
			\Line(328,-142)(370,-142)
			\Line(449,-142)(489,-142)
			\Line(208,-78)(208,-206)
			\end{picture}
			&
			\hspace*{0.5cm}\begin{picture}(147,162) (255,-191)
			\SetWidth{1.0}
			\SetColor{Black}
			\Arc(304,-78)(22.627,135,495)
			\Arc(304,-142)(22.627,135,495)
			\Line(336,-110)(401,-110)
			\Line(336,-110)(320,-94)
			\Line(288,-62)(256,-30)
			\Line(336,-110)(320,-126)
			\Line(288,-158)(256,-190)
			\end{picture}
			&
			\hspace*{2cm}\begin{picture}(225,130) (241,-207)
			\SetWidth{1.0}
			\SetColor{Black}
			\Vertex(278,-142){10}
			\Vertex(319,-142){10}
			\Vertex(399,-142){10}
			\Line(278,-142)(319,-142)
			\Line(399,-142)(465,-142)
			\Arc(359,-142)(39.598,135,495)
			\Text(442,-105)[]{\Huge{\Black{$(\tilde{Z}^{-1}_{\phi})^{T}$}}}
			\Line(278,-78)(278,-206)
			\Line(416.002,-157.998)(447.998,-126.002)\Line(416.002,-126.002)(447.998,-157.998)
			\end{picture}
			&
			\hspace*{1cm}\begin{picture}(199,130) (241,-207)
			\SetWidth{1.0}
			\SetColor{Black}
			\Line(278,-142)(439,-142)
			\Vertex(278,-142){10}
			\Text(378,-105)[]{\Huge{\Black{$(\tilde{Z}^{-1}_{\phi})^{T}$}}}
			\Line(278,-78)(278,-206)
			\Line(352.002,-157.998)(383.998,-126.002)\Line(352.002,-126.002)(383.998,-157.998)
			\end{picture}
			\\
			{\Huge $G^{(2)}_{1}$}
			&
			{\Huge $G^{(2)}_{2}$}
			&
			{\Huge $G^{(2)}_{3}$}
			&
			{\Huge $G^{(2)}_{4}$}
			&
			\hspace*{3cm}{\Huge $G^{(2)}_{5}$}
		\end{tabular*}
	}
	\caption{One- and two-loop renormalization quantities for $\phi^{3}$ theory}
	\label{12loop6d}	
\end{table}

\begin{align}
\tG^{(1)}_1&=g_1t_1, &\tG^{(1)}_2&=2\tz_1t_1, &G^{(1)}_1&=g_1t_{1g}, \nn
G^{(1)}_2&=(z_1+\tz_1)t_{1g},  & \tG^{(2)}_1&=g_2t_2,  & \tG^{(2)}_2&=g_1^2t_1^2,\nn
\tG^{(2)}_3&=2(z_1+\tz_1)g_1t_1^2+4z_1g_1t_2, & \tG^{(2)}_4&=2\tZ_2,& G^{(2)}_1&=g_2t_{2g},\nn
 G^{(2)}_2&=g_1^2(t_1^2)_g,& G^{(2)}_3&=g_1^2t_3,&&\nn
G^{(2)}_4&=(\tz_1+3z_1)(t_1^2)_g+4z_1g_1t_{2g},&
  G^{(2)}_5&=\tZ_{2g}+Z_{2g},&&
\label{gdefb}
\end{align}
where
\begin{align}
g_1&=\frac{\gamma_{11}}{\epsilon}+\gamma_{10}+\gamma_1'\epsilon+\ldots,\nn
g_2&=\frac{\gamma_{22}}{\epsilon^2}+\frac{\gamma_{11}}{\epsilon}+\gamma_{20}+\ldots,
\label{gdefa}
\end{align}
(ignoring $O(\epsilon^2)$ terms at one loop and $O(\epsilon)$ terms at two loops) and
\be
t_1^{ij}=g^{ikl}g^{jkl},\quad
t_2^{ij}=g^{ikm}g^{jlm}t_1^{kl},\quad
(t_3)^{ijk}=(t_1)^{il}(t_1)^{jm}g^{lmk}+\hbox{perms},
\ee
Furthermore, for two-index quantities such as $t_1$ we define a corresponding three-index quantity $t_{1g}$ by 
\be
t_{1g}^{ijk}=g^{ijl}(t_1)^{lk}+\hbox{perms}.
\label{tgdef}
\ee
Since (in contrast to the previous section) the two-loop 1PI diagram $\tG^{(2)}_1$ has double poles, it has a one-loop subdivergence which requires a counterterm contribution from $\tG^{(2)}_3$. There is another two-loop 2-point diagram corresponding to the tensor structure $g^{ikl}g^{jmn}g^{kmp}g^{lnp}$, but this only has a simple pole. It may therefore be dealt with in a similar fashion to the diagrams of Sect 2 and we shall return to consider it later. We are also neglecting here one and two-loop 1PI graphs giving contributions to $\hat g$ in Eq.~\eqref{gbdef}.
We now define in Eq.~\eqref{zphi}
\be
Z_1=z_1t_1,\quad Z_2=z_2t_2+x_2t_1^2, 
\label{zexpa}
\ee
with corresponding expressions in \eqref{zphi} for $\tZ_1$, $\tZ_2$.
Once again allowing for a non-minimal subtraction scheme, we write in Eq.~\eqref{zexpa} (and its analogue for $\tZ_1$, $\tZ_2$)
\begin{align}
z_1&=\frac{\zeta_{11}}{\epsilon}+\zeta_{10},\quad \tz_1=\frac{\zeta_{11}}{\epsilon}+\tzeta_{10},\nn
z_2&=\frac{\zeta_{22}}{\epsilon^2}+\frac{\zeta_{21}}{\epsilon}+\zeta_{20},\quad \tz_2=\frac{\zeta_{22}}{\epsilon^2}+\frac{\zeta_{21}}{\epsilon}+\tzeta_{20},\nn
x_2&=\tfrac32(z_1^2-\zeta_{10}^2)+\xi_4,\quad \tx_2=\tfrac32(z_1^2-\zeta_{10}^2)+(z_1-\zeta_{10})\delta_1+\txi_4,
\label{zexpb}
\end{align}
where
\be
\tzeta_{10}=\zeta_{10}+\delta_1, \quad
\tzeta_{20}=\zeta_{20}+\delta_2.
\label{zexpx}
\ee
We then require that
\begin{align}
\tG_1^{(1)}+\tG_2^{(1)}&=\hbox{finite},&\quad G_1^{(1)}+G_2^{(1)}&=\hbox{finite},\nn 
\tG_1^{(2)}+\tG_2^{(2)}+\tG_3^{(2)}+\tG_4^{(2)}&=\hbox{finite}, &\quad
G_1^{(2)}+G_2^{(2)}+G_3^{(2)}+G_4^{(2)}&=\hbox{finite}.
\end{align}
Due to the double poles in $g_2$ in Eq.~\eqref{gdefa}, the relation between the poles in $z_2$ and $\tz_2$, and those
 in $g_2$, is now non-trivial. We have as before
\be
\zeta_{11}=-\tfrac12\gamma_{11}.
\label{zexpy}
\ee
However, now we find
\begin{align}
\zeta_{22}&=\tfrac12\gamma_{22}=2\zeta_{11}^2,\nn
\zeta_{21}&=-\tfrac12(\gamma_{21}-2\gamma_{11}\gamma_{10}
+4\gamma_{11}\zeta_{10}).
\label{zgam}
\end{align}
 As in the previous section, we refrain from inserting any explicit values for the present.
Up to this order we have in Eq.~\eqref{gbdef}
\be
\hat g^{ijk}_B=x_3t_3+\ldots,
\label{xfour}
\ee
where $t_3$ was defined in Eq.~\eqref{tgdef} and
\be
x_3=z_1^2-\zeta_{10}^2+\xi_3,
\label{gfourdef}
\ee
and the ellipsis indicates the 1PI contributions.
 The MOM prescription would entail taking in Eq.~\eqref{zexpb}, \eqref{gfourdef}
\be
\zeta_{10}=-\tfrac12\gamma_{10}, \quad \zeta_{20}=-\tfrac12(\gamma_{20}-2\gamma_{10}^2-2\gamma_{11}\gamma'_1),\quad \xi_3=\zeta_{10}^2,\quad \xi_4=\tfrac32\zeta_{10}^2,
\label{momvals}
\ee
where of course $\zeta_{20}$ incorporates finite contributions from one-loop counterterms, and with the standard implementation we would also take $\tzeta_{20}=\zeta_{20}$, i.e. $\delta_2=0$.
We may now use the results from the Appendix to compute the one and two loop $\beta$-functions, just as we did in Section 2. We may again write
\be
\beta=-\tfrac12g\epsilon+(b_1+b_{01}\epsilon)
t_{1g}+(b_2+b_{02}\epsilon)t_{2g}+\ldots
\label{betrestwob}
\ee
where now we derive from Eqs.~\eqref{betares}, \eqref{Zresa} that
\be
b_1=\zeta_{11},\quad b_{01}=\zeta_{01},\quad
b_2=2(\zeta_{21}-4\zeta_{11}\zeta_{10})
=-\tfrac12(\gamma_{21}-2\gamma_{11}\gamma_{10}),\quad b_{02}=2\zeta_{20}.
\label{betrestwoa}
\ee
We may verify using Eq.~\eqref{betares} (as we did for 1PR contributions in Sect. 2) that there are no 1PR $\beta$-function contributions from the $x_3$ term in Eq.~\eqref{xfour} or from $(t_1^2)^{i(l}g^{jk)l}$. Furthermore, we see again that the $\beta$-function coefficient $b_2$ is  scheme-independent since it  may be expressed purely in terms of poles in Feynman diagrams with no dependence on our choice of finite parts such as $\zeta_{10}$, $\zeta_{20}$. This is once again consistent with the deduction from coupling redefinitions. From Eqs.~\eqref{Zresa}, \eqref{redexp},  we expect that in general the change of scheme from $\MSbar$ to a general non-minimal scheme may be effected up to two loops by
\be
\delta g=-\zeta_{10}t_{1g}.
\label{yredef}
\ee%
It is easy to check using Eqs.~\eqref{betrestwoa}, \eqref{redeflow},  that at two loops
$\delta\beta^i$ does not contain any $t_{g2}$ terms,
in agreement with the scheme independence of $b_2$ in Eq.~\eqref{betrestwoa}.

As in the previous Section, we now turn to three loops and compare a direct computation of the 1PR2 diagrams within the non-minimal scheme with the expected results from coupling redefinition.  The calculation follows the same steps as in Sect 2. Once again, we start with the three-loop 1PR2 diagrams, now as shown in Table \ref{3loop6d2}, which
give

\begin{table}[t]
	\setlength{\extrarowheight}{1cm}
	\setlength{\tabcolsep}{24pt}
	\hspace*{-9.5cm}
	\centering
	\resizebox{6.5cm}{!}{
		\begin{tabular*}{20cm}{cccc}
			\begin{picture}(283,130) (207,-207)
			\SetWidth{1.0}
			\SetColor{Black}
			\Arc(288,-142)(39.598,135,495)
			\Line(208,-142)(248,-142)
			\Line(328,-142)(369,-142)
			\Line(449,-142)(489,-142)
			\Arc(409,-142)(39.598,135,495)
			\Arc(409,-92.857)(39.143,-140.034,-39.966)
			\SetColor{White}
			\Line(208,-78)(208,-206)
			\end{picture}
			&
			\begin{picture}(283,130) (207,-207)
			\SetWidth{1.0}
			\SetColor{Black}
			\Arc(288,-142)(39.598,135,495)
			\Arc(410,-142)(39.598,135,495)
			\Line(208,-142)(248,-142)
			\Line(328,-142)(370,-142)
			\Line(449,-142)(489,-142)
			\Arc(288,-92.857)(39.143,-140.034,-39.966)
			\SetColor{White}
			\Line(208,-78)(208,-206)
			\end{picture}
			&
			\begin{picture}(187,130) (264,-207)
			\SetWidth{1.0}
			\SetColor{Black}
			\Vertex(319,-142){10}
			\Vertex(399,-142){10}
			\Line(265,-142)(319,-142)
			\Line(399,-142)(450,-142)
			\Arc(359,-142)(39.598,135,495)
			\Arc(359,-92.857)(39.143,-140.034,-39.966)
			\Line(272.002,-157.998)(303.998,-126.002)\Line(272.002,-126.002)(303.998,-157.998)
			\Line(416.002,-157.998)(447.998,-126.002)\Line(416.002,-126.002)(447.998,-157.998)
			\Text(292,-110)[]{\Huge{\Black{$\tilde{Z}^{-1}_{\phi}$}}}
			\Text(440,-110)[]{\Huge{\Black{$(\tilde{Z}^{-1}_{\phi})^{T}$}}}
			\SetColor{White}
			\Line(265,-78)(265,-206)
			\end{picture}
			&
			\begin{picture}(211,130) (255,-207)
			\SetWidth{1.0}
			\SetColor{Black}
			\Vertex(319,-142){10}
			\Vertex(399,-142){10}
			\Line(256,-142)(319,-142)
			\Line(399,-142)(465,-142)
			\Arc(359,-142)(39.598,135,495)
			\Text(288,-115)[]{\Huge{\Black{$\tilde{Z}^{-1}_{\phi}$}}}
			\Text(440,-115)[]{\Huge{\Black{$(\tilde{Z}^{-1}_{\phi})^{T}$}}}
			\Line(416.002,-157.998)(447.998,-126.002)\Line(416.002,-126.002)(447.998,-157.998)
			\Line(272.002,-157.998)(303.998,-126.002)\Line(272.002,-126.002)(303.998,-157.998)
			\SetColor{White}
			\Line(256,-78)(256,-206)
			\end{picture}
			\\
			{\Huge $\tilde{G}_{1}$}
			&
			{\Huge $\tilde{G}_{2}$}
			&
			{\Huge $\tilde{G}_{3}$}
			&
			{\Huge $\tilde{G}_{4}$}
			\\
			&
			&
			&
			\\
			\begin{picture}(189,130) (277,-207)
			\SetWidth{1.0}
			\SetColor{Black}
			\Line(278,-142)(319,-142)
			\Line(399,-142)(440,-142)
			\Arc(359,-142)(39.598,135,495)
			\CTri(380.444,-142)(396,-126.444)(411.556,-142){Black}{Black}\CTri(380.444,-142)(396,-157.556)(411.556,-142){Black}{Black}
			\CTri(307.444,-142)(323,-126.444)(338.556,-142){Black}{Black}\CTri(307.444,-142)(323,-157.556)(338.556,-142){Black}{Black}
			\SetColor{White}
			\Line(278,-78)(278,-206)
			\end{picture}
			&
			\begin{picture}(211,130) (255,-207)
			\SetWidth{1.0}
			\SetColor{Black}
			\Text(365,-70)[]{\Huge{\Black{$2$ x $one\,loop$}}}
			\Vertex(319,-142){10}
			\Vertex(399,-142){10}
			\Line(256,-142)(319,-142)
			\Line(399,-142)(465,-142)
			\Arc(359,-142)(39.598,135,495)
			\Text(288,-115)[]{\Huge{\Black{$\tilde{Z}^{-1}_{\phi}$}}}
			\Text(440,-115)[]{\Huge{\Black{$(\tilde{Z}^{-1}_{\phi})^{T}$}}}
			\Line(416.002,-157.998)(447.998,-126.002)\Line(416.002,-126.002)(447.998,-157.998)
			\Line(272.002,-157.998)(303.998,-126.002)\Line(272.002,-126.002)(303.998,-157.998)
			\SetColor{White}
			\Line(256,-78)(256,-206)
			\end{picture}
			&
			\begin{picture}(353,130) (172,-207)
			\SetWidth{1.0}
			\SetColor{Black}
			\Vertex(248,-142){10}
			\Vertex(328,-142){10}
			\Vertex(370,-142){10}
			\Vertex(449,-142){10}
			\Arc(288,-142)(39.598,135,495)
			\Arc(410,-142)(39.598,135,495)
			\Line(208,-142)(248,-142)
			\Line(328,-142)(370,-142)
			\Line(449,-142)(489,-142)
			\SetColor{White}
			\Line(208,-78)(208,-206)
			\end{picture}
			&
			\hspace*{-2cm}\begin{picture}(199,130) (241,-207)
			\SetWidth{1.0}
			\SetColor{Black}
			\Line(278,-142)(439,-142)
			\Text(338,-110)[]{\Huge{\Black{$\tilde{Z}^{-1}_{\phi}$}}}
			\Text(408,-110)[]{\Huge{\Black{$(\tilde{Z}^{-1}_{\phi})^{T}$}}}
			\Line(352.002,-157.998)(383.998,-126.002)\Line(352.002,-126.002)(383.998,-157.998)
			\SetColor{White}
			\Line(278,-78)(278,-206)
			\end{picture}
			\\
			{\Huge $\tilde{G}_{5}$}
			&
			{\Huge $\tilde{G}_{6}$}
			&
			{\Huge $\tilde{G}_{7}$}
			&
			{\Huge $\tilde{G}_{8}$}
		\end{tabular*}
	}
	\caption{Three-loop, two-point renormalization quantities for $\phi^{3}$ theory}
	\label{3loop6d2}	
\end{table}

\begin{align}
\tG^{(3)}_{1}&=g_1g_2t_1t_2,\nn
\tG^{(3)}_{2}&=g_1g_2t_2t_1\nn
\tG^{(3)}_3&=g_2(z_1+\tz_1)(t_1t_2+t_2t_1),\nn
\tG^{(3)}_4&=g_1(z_2+\tz_2)(t_1t_2+t_2t_1),\nn
\tG^{(3)}_5&=2(z_1+2\tz_1)z_1g_1(t_1t_2+t_2t_1),\nn
\tG^{(3)}_6&=2z_1^2g_1(t_1t_2+t_2t_1),\nn
\tG^{(3)}_7&=4z_1g_1^2(t_1t_2+t_2t_1),\nn
\tG^{(3)}_8&=\tz_1\tz_2(t_1t_2+t_2t_1)+\tZ_3+\tZ_3^T
\label{Ddiaga}
\end{align}
where we write
\be
\tZ_3=\tz_3t_1t_2+\tz_3't_2t_1+\ldots,
\label{zdefe}
\ee
with
\be
\tz_3=\frac{\tzeta_{33}}{\epsilon^3}+\frac{\tzeta_{32}}{\epsilon^2}+\frac{\tzeta_{31}}{\epsilon}
+\tzeta_{30},\quad
\tz_3'=\frac{\tzeta'_{33}}{\epsilon^3}+\frac{\tzeta'_{32}}{\epsilon^2}+\frac{\tzeta'_{31}}{\epsilon}
+\tzeta'_{30},
\label{zdefd}
\ee
and
\be
\tzeta_{30}=\zeta_{30}+\delta_{30},\quad
\tzeta'_{30}=\zeta'_{30}+\delta'_{30},
\ee
where $\zeta_{30}$, $\zeta'_{30}$ are quantities corresponding to  $\tzeta_{30}$, $\tzeta'_{30}$ in $Z_3$, defined later in Eqs.~\eqref{zdeff}, \eqref{zdefg}.
We again neglect here the 1PI three-loop two-point diagrams.

We shall see that the difference between $Z$ and $\tZ$ at two loops due to the presence of $\delta_1$, $\delta_2$ implies that the coefficients of the simple {\it and double} poles in $t_2t_1$ and $t_1t_2$ in $Z_3$ must differ from those in $\tZ_3$. This is a difference from Sect 2 where only the simple pole terms in $T_2T_1$ and $T_1T_2$ were different (due of course to the double pole now present in $Z_2$). The other pole terms at three loops are unaffected, i.e are the same as for $\delta_1=\delta_2=0$.

We have
\be
\sum \tG_{\alpha}^{(3)}=\hbox{finite}.
\ee
Again, as we saw in Section 2, this condition only defines sums such as $\tzeta_{31}+\tzeta_{31}'$; but now we shall find that we need to allow for
$\tzeta_{32}$, $\tzeta_{32}'$ to be different, as well as $\tzeta_{31}$, $\tzeta_{31}'$. We find, inserting Eqs.~\eqref{gdefa}, \eqref{zexpa}-\eqref{zexpx}, \eqref{zexpy}, \eqref{zgam} into Eq.~\eqref{Ddiaga},
\begin{align}
\tzeta_{33}&=\tzeta_{33}'=\tfrac32\zeta_{22}\zeta_{11},\nn
\tzeta_{32}+\tzeta_{32}'&=3C_2+2\zeta_{11}^2\delta_1,\nn
\tzeta_{31}+\tzeta'_{31}&=3C_1+\zeta_{11}(\delta_2+4\xi_3-4\zeta_{10}^2)
+\zeta_{21}\delta_1,
\label{zzprimea}
\end{align}
where now
\begin{align}
C_1&=\zeta_{11}\zeta_{20}+\zeta_{21}\zeta_{10},\nn
C_2&=\zeta_{21}\zeta_{11}+\zeta_{22}\zeta_{10}.
\label{Cdefs}
\end{align}
We write
\begin{align}
\tzeta_{32}&=\tfrac32C_2+\zeta_{11}^2\delta_1+\xi_2,\nn \tzeta_{32}'&=\tfrac32C_2+\zeta_{11}^2\delta_1-\xi_2,\nn
\tzeta_{31}&=\tfrac32C_1+\tfrac12(\zeta_{11}\delta_2
+\zeta_{21}\delta_1)+2\zeta_{11}(\xi_3-\zeta_{10}^2)+\xi_1,\nn
\tzeta'_{31}&=\tfrac32C_1+\tfrac12(\zeta_{11}\delta_2
+\zeta_{21}\delta_1)+2\zeta_{11}(\xi_3-\zeta_{10}^2)-\xi_1.
\label{zzprimeb}
\end{align}
We now start to see how despite the added complexity of the intermediate calculations, the final results mirror those obtained earlier in Section 2.

We now again use our indirect method  which will enable us to reconstruct $\xi_1$  from the asymmetric part of the anomalous dimension as obtained by coupling redefinition. The result for $\xi_2$ will also be obtained as a by-product.This enables us to deduce the individual simple pole coefficients $\tzeta_{31}$, $\tzeta_{31}'$, and double pole coefficients $\tzeta_{32}$, $\tzeta_{32}'$. We write
\be
\gamma_{\phi}=\gamma_1t_1+\gamma_2t_2+\gamma_{3}t_1t_2+\gamma'_{3}t_2t_1+\ldots
\ee
where the ellipsis indicates irrelevant three-loop terms as well as higher-order contributions. Inserting \eqref{zdefe} with the results of Eq.~\eqref{Zresa}  into Eq.~\eqref{gamthree} we find
\begin{align}
    \gamma_{\phi}^{(3)}&=\epsilon(3\tz_3-\tz_1\tz_2)-(4\tx_2+\tz_2)(b_1+b_{01}\epsilon)-\tz_1(b_2+b_{02}\epsilon]t_1t_2\nn
&+[\epsilon(3\tz'_3-2\tz_1\tz_2)+(4\tz_1^2-4\tx_2-\tz_2)(b_1+b_{01}\epsilon)-\tz_1(b_2
    +b_{02}\epsilon)]t_2t_1+\ldots,
\end{align}
where the $\beta$-function coefficients are listed in Eq.~\eqref{betrestwoa}.
Imposing finiteness on the right-hand side and using Eqs.~\eqref{zexpb}, \eqref{zdefd} yields $\tzeta_{33}$ in agreement with Eq.~\eqref{zzprimea}, and
\be
\tzeta_{32}=\tfrac43C_2+2\zeta_{11}^2\zeta_{10}+2\zeta_{11}^2\delta_1,\quad
\tzeta'_{32}=\tfrac53C_2-2\zeta_{11}^2\zeta_{10},
\label{xitwo}
\ee
together with
\begin{align}
\gamma_3&=3\tzeta_{31}-\zeta_{11}(2\zeta_{20}+2\tzeta_{20})-\zeta_{21}
(\zeta_{10}+3\tzeta_{10})-2\zeta_{11}(4\zeta_{10}^2-2\zeta_{10}\tzeta_{10}+2\txi_4),\nn
\gamma'_3&=3\tzeta'_{31}-\zeta_{11}(2\zeta_{20}+3\tzeta_{20})-\zeta_{21}
(\zeta_{10}+4\tzeta_{10})-2\zeta_{11}(4\zeta_{10}^2-6\zeta_{10}\tzeta_{10}-2\tzeta_{10}^2+2\txi_4).
\label{gammathree}
\end{align}
Eq.~\eqref{xitwo} is consistent with Eq.~\eqref{zzprimea}, and furthermore comparison with Eq.~\eqref{zzprimeb} yields
\be
\xi_2=-\tfrac16C_2+2\zeta_{11}^2\zeta_{10}+\zeta_{11}^2\delta_1.
\label{xitwoa}
\ee
As in  Section 2, the difference $\gamma_3-\gamma'_3$ may also be computed by making the appropriate coupling and field redefinition from $\MSbar$. The field redefinition is given by $\phi'=\Omega\phi$  where as we see from Eq.~\eqref{redexp}
\be
\delta \Omega =\omega_{1}t_1+\omega_2t_2
\label{omdef}
\ee
where $\omega_1=\tzeta_{10}$, $\omega_2=\tzeta_{20}-4\zeta_{10}\tzeta_{10}$. Now from the symmetry of $\gamma_{\phi}$ in $\MSbar$, we deduce from Eq.~\eqref{gamredgen} in Appendix B 
\be
\gamma_3-\gamma_3'=2(\gamma_2\omega_{1}-\gamma_{1}\omega_2)-4\omega_1^2\gamma_1
\label{gammtwoa}
\ee
where $\gamma_1=\zeta_{11}$, $\gamma_{2}=2\zeta_{21}-8\zeta_{10}\zeta_{11}$ (note that we had to include terms 2nd order in $\delta\Omega$ here).
Finally we deduce from Eqs.~\eqref{zzprimeb},  \eqref{gammathree} and \eqref{gammtwoa} that
\be
\xi_1=\tfrac12(\tzeta_{31}-\tzeta'_{31})=\tfrac12(\zeta_{21}\tzeta_{10}-\zeta_{11}\tzeta_{20}),
\label{xiresa}
\ee
i.e. the same form as in Sect. 2.
Inserting this expression for $\xi_1$ into Eq.~\eqref{zzprimeb} gives 
\begin{align}
\tzeta_{31}&=C_1+\zeta_{21}\tzeta_{10}+2\zeta_{11}(\xi_3-\zeta_{10}^2),\nn
\tzeta'_{31}&=C_1+\zeta_{11}\tzeta_{20}+2\zeta_{11}(\xi_3-\zeta_{10}^2).
\label{zzprimenn}
\end{align}
As in Sect 2, we now consider the 1PR three point diagrams as depicted in Table~\ref{3loop6d3}, corresponding to the 1PR 2-point diagrams of Table \ref{3loop6d2}.
These diagrams give contributions to the 3-point function given by

\begin{table}[t]
	\setlength{\extrarowheight}{1cm}
	\setlength{\tabcolsep}{24pt}
	\hspace*{-8.5cm}
	\centering
	\resizebox{6.5cm}{!}{
		\begin{tabular*}{20cm}{cccc}
			\begin{picture}(283,130) (207,-207)
			\SetWidth{1.0}
			\SetColor{Black}
			\Arc(288,-142)(39.598,135,495)
			\Line(208,-142)(248,-142)
			\Line(328,-142)(369,-142)
			\Line(449,-142)(489,-142)
			\Arc(409,-142)(39.598,135,495)
			\Arc(409,-92.857)(39.143,-140.034,-39.966)
			\Line(208,-78)(208,-206)
			\end{picture}
			&
			\begin{picture}(283,130) (207,-207)
			\SetWidth{1.0}
			\SetColor{Black}
			\Arc(288,-142)(39.598,135,495)
			\Arc(410,-142)(39.598,135,495)
			\Line(208,-142)(248,-142)
			\Line(328,-142)(370,-142)
			\Line(449,-142)(489,-142)
			\Arc(288,-92.857)(39.143,-140.034,-39.966)
			\Line(208,-78)(208,-206)
			\end{picture}
			&
			\begin{picture}(211,130) (255,-207)
			\SetWidth{1.0}
			\SetColor{Black}
			\Vertex(256,-142){10}
			\Vertex(319,-142){10}
			\Vertex(399,-142){10}
			\Line(256,-142)(319,-142)
			\Line(399,-142)(465,-142)
			\Arc(359,-142)(39.598,135,495)
			\Text(440,-105)[]{\Huge{\Black{$(\tilde{Z}^{-1}_{\phi})^{T}$}}}
			\Line(416.002,-157.998)(447.998,-126.002)\Line(416.002,-126.002)(447.998,-157.998)
			\Line(256,-78)(256,-206)
			\end{picture}
			&
			\begin{picture}(187,130) (264,-207)
			\SetWidth{1.0}
			\SetColor{Black}
			\Vertex(265,-142){10}
			\Vertex(319,-142){10}
			\Vertex(399,-142){10}
			\Line(265,-142)(319,-142)
			\Line(399,-142)(450,-142)
			\Arc(359,-142)(39.598,135,495)
			\Arc(359,-92.857)(39.143,-140.034,-39.966)
			\Line(416.002,-157.998)(447.998,-126.002)\Line(416.002,-126.002)(447.998,-157.998)
			\Text(440,-105)[]{\Huge{\Black{$(\tilde{Z}^{-1}_{\phi})^{T}$}}}
			\Line(265,-78)(265,-206)
			\end{picture}
			\\
			{\Huge $G_{1}$}
			&
			{\Huge $G_{2}$}
			&
			{\Huge $G_{3}$}
			&
			{\Huge $G_{4}$}
			\\
			&
			&
			&
			\\
			\begin{picture}(189,130) (277,-207)
			\SetWidth{1.0}
			\SetColor{Black}
			\Line(278,-142)(319,-142)
			\Line(399,-142)(440,-142)
			\Arc(359,-142)(39.598,135,495)
			\CTri(380.444,-142)(396,-126.444)(411.556,-142){Black}{Black}\CTri(380.444,-142)(396,-157.556)(411.556,-142){Black}{Black}
			\CTri(307.444,-142)(323,-126.444)(338.556,-142){Black}{Black}\CTri(307.444,-142)(323,-157.556)(338.556,-142){Black}{Black}
			\Line(278,-78)(278,-206)
			\end{picture}
			&
			\begin{picture}(211,130) (255,-207)
			\SetWidth{1.0}
			\SetColor{Black}
			\Vertex(256,-142){10}
			\Vertex(319,-142){10}
			\Vertex(399,-142){10}
			\Text(365,-60)[]{\Huge{\Black{$2$ x $one\,loop$}}}
			\Line(256,-142)(319,-142)
			\Line(399,-142)(465,-142)
			\Arc(359,-142)(39.598,135,495)
			\Text(440,-105)[]{\Huge{\Black{$(\tilde{Z}^{-1}_{\phi})^{T}$}}}
			\Line(416.002,-157.998)(447.998,-126.002)\Line(416.002,-126.002)(447.998,-157.998)
			\Line(256,-78)(256,-206)
			\end{picture}
			&
			\begin{picture}(353,130) (172,-207)
			\SetWidth{1.0}
			\SetColor{Black}
			\Vertex(248,-142){10}
			\Vertex(328,-142){10}
			\Vertex(370,-142){10}
			\Vertex(449,-142){10}
			\Arc(288,-142)(39.598,135,495)
			\Arc(410,-142)(39.598,135,495)
			\Line(208,-142)(248,-142)
			\Line(328,-142)(370,-142)
			\Line(449,-142)(489,-142)
			\Line(208,-78)(208,-206)
			\end{picture}
			&
			\hspace*{-2cm}\begin{picture}(199,130) (241,-207)
			\SetWidth{1.0}
			\SetColor{Black}
			\Vertex(278,-142){10}
			\Line(278,-142)(439,-142)
			\Text(375,-105)[]{\Huge{\Black{$(\tilde{Z}^{-1}_{\phi})^{T}$}}}
			\Line(352.002,-157.998)(383.998,-126.002)\Line(352.002,-126.002)(383.998,-157.998)
			\Line(278,-78)(278,-206)
			\end{picture}
			\\
			{\Huge $G_{5}$}
			&
			{\Huge $G_{6}$}
			&
			{\Huge $G_{7}$}
			&
			{\Huge $G_{8}$}
		\end{tabular*}
	}
	\caption{Three-loop, three-point renormalization quantities for $\phi^{3}$ theory}
	\label{3loop6d3}	
\end{table}

\begin{align}
G^{(3)}_{1}&=g_1g_2(t_1t_2)_g,\nn
G^{(3)}_{2}&=g_1g_2(t_2t_1)_g,\nn
G^{(3)}_3&=2g_2z_1(t_1t_2)_g+g_2(z_1+\tz_1)(t_2t_1)_g,\nn
G^{(3)}_4&=2g_1z_2(t_2t_1)_g+g_1(z_2+\tz_2)(t_1t_2)_g,\nn
G^{(3)}_5&=2g_1\tz_1[(\tz_1+2z_1)(t_2t_1)_g+3\tz_1(t_1t_2)_g],\nn
G^{(3)}_6&=2z_1^2g_1(t_1t_2+t_2t_1)_g,\nn
G^{(3)}_7&=4g_1^2z_1(t_1t_2+t_2t_1)_g,\nn
G^{(3)}_8&=z_1\tz_2(t_1t_2)_g+z_2\tz_1(t_2t_1)_g+Z_{3g}+\tZ_{3g}^T
\label{Ddiagb}
\end{align}
We have
\be
\sum G^{(3)}_{\alpha}=\hbox{finite}.
\ee
We write, analogously to Eq.~\eqref{zdefe},
\be
Z_3=z_3t_1t_2+z_3't_2t_1+\ldots
\label{zdeff}
\ee
with
\be
z_3=\frac{\zeta_{33}}{\epsilon^3}+\frac{\zeta_{32}}{\epsilon^2}+\frac{\zeta_{31}}{\epsilon}
+\zeta_{30},\quad
z_3'=\frac{\zeta_{33}}{\epsilon^3}+\frac{\zeta'_{32}}{\epsilon^2}+\frac{\zeta'_{31}}{\epsilon}
+\zeta'_{30}.
\label{zdefg}
\ee
Inserting Eqs.~\eqref{gdefa}, \eqref{zexpa}-\eqref{zexpx}, \eqref{zexpy}, \eqref{zgam} and \eqref{zzprimeb} into Eq.\eqref{Ddiagb}, we obtain
\begin{align}
\zeta_{33}&=\zeta_{33}'=\tfrac32\zeta_{22}\zeta_{11},\nn
\zeta_{32}&=\tfrac32C_2-\zeta_{11}^2\delta_1+\xi_2, \quad\zeta_{32}'=\tfrac32C_2+\zeta_{11}^2\delta_1-\xi_2,\nn
\zeta_{31}&=\tfrac32C_1+2\zeta_{11}(\xi_3-\zeta_{10}^2)
+\tfrac12(\zeta_{11}\delta_2-\zeta_{21}\delta_1)+\xi_1,\nn
\zeta'_{31}&=\tfrac32C_1+2\zeta_{11}(\xi_3-\zeta_{10}^2)+\tfrac12(\zeta_{21}\delta_1
-\zeta_{11}\delta_2)-\xi_1,
\end{align}
where again $\delta_1$, $\delta_2$ are defined in Eqs.~\eqref{zexpx}. We see using Eqs.~\eqref{xitwoa}, \eqref{xiresa},
\begin{align}
\tzeta_{32}&=\tfrac43C_2+2\zeta_{11}^2\zeta_{10},\nn
\tzeta'_{32}&=\tfrac53C_2-2\zeta_{11}^2\zeta_{10},\nn
\zeta_{31}&=C_1+\zeta_{21}\zeta_{10}+2\zeta_{11}(\xi_3-\zeta_{10}^2),\nn
\zeta'_{31}&=C_1+\zeta_{11}\zeta_{20}+2\zeta_{11}(\xi_3-\zeta_{10}^2).
\label{tzfina}
\end{align}
Using Eq.~\eqref{betares}, and writing $b_3$, $b_3'$ for the coefficients of $(t_1t_2)_g$
and $(t_2t_1)_g$, respectively, in the $\beta$-function, we find
\begin{align}
b_3&=3\zeta_{31}-4C_1+4(\zeta_{10}^2-\xi_4-\xi_3)\zeta_{11},\nn
b'_3&=3\zeta'_{31}-5C_1+4(4\zeta_{10}^2-\xi_4-\xi_3)\zeta_{11},
\label{betexp}
\end{align}
and combining Eqs.~\eqref{tzfina}, \eqref{betexp}, we finally obtain
\begin{align}
b_3&=-b_2\delta_1+b_1(\zeta_{20}-4\xi_4+2\xi_3+6\zeta_{10}^2),\nn
b'_3&=b_2\delta_1-b_1(\zeta_{20}+4\xi_4-2\xi_3+2\zeta_{10}^2),\nn
\zeta_{32}&=\tfrac43C_2+2\zeta_{11}^2\zeta_{10},\nn
\zeta'_{32}&=\tfrac53C_2-2\zeta_{11}^2\zeta_{10},
\label{betfina}
\end{align}
with the one and two loop $\beta$-function coefficients $b_1$, $b_2$ as given in Eq.~\eqref{betrestwoa}.
A reassuring immediate check on the calculation is that the results for $\zeta_{32}$, $\zeta'_{32}$ agree with those for
$\tzeta_{32}$, $\tzeta'_{32}$ in Eq.~\eqref{xitwo}, except that $\delta_1$ has now completely cancelled; this must be the case since these coefficients are determined in terms of lower-order $\beta$-function coefficients by Eq.~\eqref{poleres} (whereas
$\tzeta_{32}$, $\tzeta'_{32}$ were determined by Eq.~\eqref{gamthree} which also involves lower-order contributions to $\tZ_{\phi}$). A more detailed check verifies that these coefficients are indeed exactly in accord with Eq.~\eqref{poleres}. In performing this check and also in computing $b_3$, $b_3'$ in Eq.~\eqref{betexp}, it is crucial to incorporate the $x_2$ terms in Eq.~\eqref{zexpa}.

We now  derive the predictions for the 1PR anomalous dimension contributions for a general non-minimal scheme, as obtained by a scheme change from $\MSbar$. The change of renormalisation scheme corresponds to a coupling redefinition which can be derived from Eq.~\eqref{redexp} using Eqs.~\eqref{Zresa} (noting that we need to work to second order in $\eta_1$). We find
\be
\delta g=\eta_1t_{1g}+\eta_2t_{2g}+\eta_3t_3+\eta_4[(t_1)^2]_g,
\label{delg}
\ee%
where $t_i$ were defined in Eq.~\eqref{tgdef} and 
\be
\eta_1=-\zeta_{10},\quad \eta_2=-\zeta_{20}+4\zeta_{10}^2,\quad \eta_3=-\xi_3+2\zeta_{10}^2,\quad \eta_4=-\xi_4+3\zeta_{10}^2.
\label{etadef}
\ee
We use the general results for the effect of a scheme change given in Eq.~\eqref{redeffull} in conjunction with Eqs.~\eqref{betrestwob}, \eqref{betrestwoa} (again working to second order in $\eta_1$), and find
\be
\delta\beta=\delta b_3(t_1t_2)_g+\delta b_3'(t_2t_1)_g+\ldots,
\ee
where
\begin{align}
\delta b_3'&=b_{2}\zeta_{10}+b_{1}(\eta_2-2\eta_3+4\eta_4-6\eta_1^2),\nn
\delta b_3&=-b_{2}\zeta_{10}+b_{1}(-\eta_2-2\eta_3+4\eta_4-2\eta_1^2),
\label{deltwoa}
\end{align}
with $b_1$, $b_2$ as in Eq.~\eqref{betrestwoa}.
Since in $\MSbar$ we have $b_3^{\MSbar}=b_3^{\prime\MSbar}=0$, we find agreement between Eq.~\eqref{betfina} and Eqs.~\eqref{etadef}, \eqref{deltwoa}.

\section{Conclusions}
We have shown that starting at three loop order the $\beta$-function derived using a non-minimal renormalisation scheme contains terms corresponding to 1PR contributions to the two-point function, using as examples the cases of theories with general couplings in both four and six dimensions. We were unable to perform a full explicit three-loop computation of the relevant terms, since the computation  of the two-point function yields only the symmetrised part. We therefore relied on a scheme redefinition from $\MSbar$ to provide further information on the asymmetric part. Therefore the most we can say is that the scheme redefinition results are consistent with those obtained explicitly. 

We have considered a general non-minimal scheme in which the two-point function is renormalised differently from the three point function; i.e. 1PR three point diagrams with divergent two-point subdiagrams are assigned different finite parts (via $Z_{\phi}$) to those appearing in the corresponding two-point renormalisation constant for $\tZ_{\phi}$; this is a somewhat different philosophy from the ``hybrid MOM'' schemes\cite{smekal}, where $Z_{\phi}=\tZ_{\phi}$ but the 1PI three-point diagrams are renormalised by a different prescription. In both the four-dimensional and six-dimensional case we have only carried out a partial computation of the 1PR2 terms in which we are interested; in four dimensions we focussed on mixed fermion scalar terms for simplicity, since in this case the relevant two-loop contribution to $\tZ_{\phi}$ had only a simple pole; and in six dimensions we omitted a two-loop simple pole contribution to $\tZ_{\phi}$ precisely so that we coould focus on the double pole contribution which displayed new features. It is now clear that we easily obtain the general results simply by incorporating extra finite contributions analogous to $\zeta_{10}$ and $\zeta_{20}$ for each relevant one and two loop tensor structure; in the four-dimensional case this would involve consideration of  $\tZ_{\psi}$ in addition to $\tZ_{\phi}$.

 An interesting special case is where we set $\tzeta_{10}=\tzeta_{20}=0$, i.e. renormalise the two-point function minimally (note that this is not $\MSbar$, since the counterterms are those computed in the non-minimal scheme).In this case we have $\delta\Omega=0$ and no field redefinition is required; we also find $\xi=0$ in Section 2 and $\xi_1=0$ in Section 3, indicating that in this case the contributions to $\tZ$ are predicted to be symmetric.

  Our own interest in this issue was inspired though the consistency conditions on $\beta$-function coefficients, derived from gradient flow equations which were obtained during the investigation of a six-dimensional $a$-theorem\cite{JP1}. Specifically, the appearance of potential 1PR $\beta$-function contributions was noticed during the demonstration of the scheme independence of these consisteny conditions. There are still unresolved issues here connected with the scheme dependence of extra terms arising in the conformal anomaly for theories with an $O(n)$ symmetry, and it would be interesting to return to this with our better understanding of the general issue of scheme dependence, particularly the role of the field redefinition in this case.

\bigskip

\bigskip

{\Large{{\bf{Acknowledgements}}} }\hfil

We are very grateful to John Gracey and Tim Jones for useful conversations.
This work was supported in part by the STFC under contract ST/G00062X/1, and CP was supported by an STFC studentship.

\appendix
\section{Explicit values}
Here for completeness we record some explicit values of quantities introduced in Section 3. The explicit values in Eq.~\eqref{gdefa} are given by
\begin{align}
\gamma_{11}=-\tfrac16,\quad \gamma_{10}&=-\tfrac29,\quad \gamma'_1=-\tfrac{13}{54},\nn
\gamma_{22}=\tfrac{1}{36},\quad \gamma_{21}&=\tfrac{43}{432},\quad \gamma_{20}=\tfrac{1207}{5184}
\end{align}
suppressing a standard factor of $(64\pi^3)^{-1}$ for each loop order.
Consequently, we have
\begin{align}
\zeta_{11}&=-\tfrac12\gamma_{11}=\tfrac{1}{12},\nn
\zeta_{22}&=\tfrac12\gamma_{22}=2\zeta_{11}^2=\tfrac{1}{72}.
\end{align}
These results are scheme independent. For interest, we also quote the scheme-dependent results for the simple-pole and finite quantities, within MOM. We have 
\begin{align}
\zeta_{10}&=-\tfrac12\gamma_{10}=\tfrac{1}{9},\nn
\zeta_{21}&=-\tfrac12(\gamma_{21}-2\gamma_{11}\gamma_{10}
+4\gamma_{11}\zeta_{10})=\tfrac{7}{288},\nn
\zeta_{20}&=-\tfrac12(\gamma_{20}-2\gamma_{10}^2-2\gamma_{11}\gamma'_1)=-\tfrac{31}{1152}.
\label{zgama}
\end{align}
These results
\section{General results}
We write %
\be%
g^A_B=\mu^{k^A\epsilon}\left(g^A+\sum_{L,m=0}^{m=L}%
\frac{g^A_{Lm}}{\epsilon^m}\right)%
\label{gb}
\ee%
where $g^A$ denotes a generic set of couplings, and similarly introduce renormalisation constants $Z^{ij}$ for the fields $\phi^i$, where
\be%
Z_{\phi}=\sum_{L,m=0}^{m=L}%
\frac{Z_{Lm}}{\epsilon^m}%
\label{zgen}
\ee%
(with similar expressions for $\tZ_{\phi}$, and for the fermion field renormalisation constants where they appear). In Eq.~\eqref{gb}, there is no sum over the index on $k^A$. We have $k=\tfrac12$ for Yukawa couplings in four dimensions and scalar ($\phi^3$) couplings in six dimensions; and $k=1$ for scalar ($\phi^4$) couplings in four dimensions.
We then define%
\be
\beta^A_g=\mu\frac{d}{d\mu} g^A,
\ee
and expand
\be%
\beta^A=-\tfrac12g^A\epsilon+\sum_L\beta^A_{0L}\epsilon%
+\sum_L\beta^A_L,%
\ee%
where of course $\beta^A_L$ and $\beta^A_{0L}$ denote the $L$-loop contributions which are $O(g^{2L+1})$.
We find
\begin{align}
\beta^A_{01}&=g^A_{10},\nn%
\beta^A_{1}&=g^A_{11},\nn%
\beta^A_{02}&=2g^A_{20}-g_{10}\cdot\frac{\pa}{\pa g}g^A_{10},\nn%
\beta^A_{2}&=2g^A_{21}-\beta_{01}\cdot\frac{\pa}{\pa g}g^A_{11}
-\beta_1\cdot\frac{\pa}{\pa g}g^A_{10},\nn%
\beta^A_3&=3g^A_{31}-\beta_{02}\cdot\frac{\pa}{\pa g}g^A_{11}
-\beta_{01}\cdot\frac{\pa}{\pa g}g^A_{21}\nn
-&\beta_{2}\cdot\frac{\pa}{\pa g}g^A_{10}
-\beta_{1}\cdot\frac{\pa}{\pa g}g^A_{20},%
\label{betares}
\end{align}%
and also the useful results for the higher-order poles
\begin{align}
2g^A_{22}&=\beta_1\cdot\frac{\pa}{\pa g}\beta^A_1,\nn
3g^A_{33}&=\beta_1\cdot\frac{\pa}{\pa g}g^A_{22},\nn
3g^A_{32}&=\beta_{01}\cdot\frac{\pa}{\pa g}g^A_{22}+\beta_1\cdot\frac{\pa}{\pa g}g^A_{21}+\beta_2\cdot\frac{\pa}{\pa g}g^A_{11}.
\label{poleres}
\end{align}
We use the shorthand notation
\be
X\cdot\frac{\pa}{\pa g}\equiv X^A\frac{\pa}{\pa g^A}.
\ee
Defining
\be
\mu\frac{d}{d \mu}\phi=-\gamma\phi,
\ee
or of course equivalently
\be%
\gamma=Z_{\phi}^{-1}\mu\frac{d}{d \mu} Z_{\phi},%
\label{gamdef}
\ee%
we obtain
\begin{align}%
\gamma_3&=-3\tZ_{31}+2(\tZ_{10}\tZ_{21}+\tZ_{11}\tZ_{20})+\tZ_{21}\tZ_{10}+\tZ_{20}\tZ_{11}\nn
&+\beta_2\cdot\frac{\pa}{\pa g}\tZ_{10}+\beta_{02}\cdot\frac{\pa}{\pa g}\tZ_{11}\nn%
&+\beta_1\cdot\frac{\pa}{\pa g}\tZ_{20}+\beta_{01}\cdot\frac{\pa}{\pa g}\tZ_{21}\nn%
&-\tZ_{10}\beta_{01}\cdot\frac{\pa}{\pa g}\tZ_{11}-\tZ_{10}\beta_{1}\cdot\frac{\pa}{\pa g}\tZ_{11}-\tZ_{11}\beta_{01}\cdot\frac{\pa}{\pa g}\tZ_{10}.
\label{gamthree}
\end{align}%
For the $d=4$ scalar fermion theory, in terms of the general notation of Eq.~\eqref{zgen}
\begin{align}
\tZ_{10}&=\tzeta_{10}T_1,\quad Z_{11}=\zeta_{11}T_1, \nn \tZ_{20}&=\tzeta_{20}T_2+\txi_4T_1^2,\nn \tZ_{21}&=\zeta_{21}T_2+\zeta_{11}(3\zeta_{10}+\delta_1)T_1^2,\nn
\tZ_{22}&=\tfrac32\zeta_{11}^2T_1^2.
\label{Zres}
\end{align}
The corresponding results for $Z_{10}$ etc may be obtained from
those for $\tZ_{10}$ etc in Eq.~\eqref{Zres} simply by setting
$\delta_1=\delta_2=0$, and $\txi_4=\xi_4$. The result for $Z_{Lm}$ leads to a contribution to $g_{Lm}$ (in the notation of Eq.~\eqref{gb}, but in this case corresponding to the Yukawa coupling) given by
$(Z_{Lm})_y$ in the notation introduced in Eq.~\eqref{Gdef}.

For the $d=6$ scalar theory, in terms of the general notation of Eq.~\eqref{zgen}
\begin{align}
\tZ_{10}&=\tzeta_{10}t_1,\quad Z_{11}=\zeta_{11}t_1, \nn
\tZ_{20}&=\tzeta_{20}t_2+\txi_4t_1^2,\nn
\tZ_{21}&=\zeta_{21}t_2+\zeta_{11}(3\zeta_{10}+\delta_1)t_1^2,\nn \tZ_{22}&=\zeta_{22}t_2+\tfrac32\zeta_{11}^2t_1^2.
\label{Zresa}
\end{align}
Once again, the corresponding results for $Z_{10}$ etc may be obtained (up to this order) from
those for $\tZ_{10}$ etc in Eq.~\eqref{Zresa} simply by setting
$\delta_1=\delta_2=0$, and $\txi_4=\xi_4$; and the result for $Z_{ij}$ leads to a contribution to $g_{ij}$ (again in the notation of Eq.~\eqref{gb}, but now corresponding to the scalar coupling) given by
$Z_g$ in the notation introduced in Eq.~\eqref{tgdef}.

It is useful to derive the general redefinition relating $\MSbar$ and a non-minimal scheme. For $\MSbar,$ Eqs.~\eqref{gb}, \eqref{zgen} take the form
\begin{align}%
g^A_B&=\mu^{k^A\epsilon}\left(g^{i\MSbar}+\sum_{L=1,m=1}^{m=L}%
\frac{g_{Lm}^{i\MSbar}}{\epsilon^m}\right),\nn%
\tZ_{\phi}&=\sum_{L,m=1}^{m=L}%
\frac{\tZ_{Lm}}{\epsilon^m},
\label{gms}
\end{align}%
i.e. with the summations starting at $m=1$ rather than $m=0$,
so that comparing the finite terms in Eqs. \eqref{gb}, \eqref{zgen} with Eq.~\eqref{gms} we have the simple relations
\begin{align}
g^{\MSbar A}&=g^A+\sum_{L=1}g_{L0}^A(g),\nn
\phi^{\MSbar}&=\phi+\sum_{L=1}\tZ_{L0}^A(g)\phi
\label{coupred}
\end{align}
However we usually wish to change scheme from $\MSbar$ to the non-minimal scheme. Solving Eq.~\eqref{coupred} for $g$, $\phi$, in terms of $g^{\MSbar}$, $\phi^{\MSbar}$, we find up to second order (which will be sufficient for our purposes)
\begin{align}
g^A&=g^{\MSbar A}-g^A_{10}(g^{\MSbar})-g^A_{20}(g^{\MSbar})+g_{10}(g^{\MSbar})\cdot\frac{d}{dg^{\MSbar}}g^A_{10}(g^{\MSbar}),\nn
\phi&=\left[1-\tZ_{10}(g^{\MSbar})-\tZ_{20}(g^{\MSbar})+g_{10}(g^{\MSbar})\cdot\frac{d}{dg^{\MSbar}}\tZ_{10}(g^{\MSbar})\right]\phi^{\MSbar}.
\label{redexp}
\end{align}

Finally, we need results for the effect of a scheme change as implemented by a redefinition of couplings and fields. Under a coupling redefinition $g'\equiv g'(g)$, we have
\be
\beta^{\prime A}(g')=\mu\frac{d}{d\mu}g^{\prime A}=\beta(g)\cdot\frac{d}{dg}g^{\prime A},
\label{redeffull}
\ee
which at lowest order gives
\be
\delta\beta^A=\beta\cdot\frac{d}{dg}\delta g^A-\delta g\cdot\frac{d}{dg}\beta^A.
\label{redeflow}
\ee
We also have (as we saw in Eq.~\eqref{redexp} in the case of a change from $\MSbar$)
\be
\phi'=\Omega\phi
\ee
for some matrix $\Omega$, 
which entails
\be
\gamma'=\Omega\gamma\Omega^{-1}-\left(\mu\frac{d}{d\mu}\Omega\right)\Omega^{-1}
\label{gamredgen}
\ee
which at lowest order gives
\be
\delta\gamma=[\delta\Omega,\gamma]+\beta\cdot\frac{d}{dg}\Omega-\delta g\cdot\frac{d}{dg}\gamma.
\label{gamredef}
\ee
The computation of the two-point function yields corrections
to $\tZ_{\phi}\tZ_{\phi}^T$ rather than to $\tZ_{\phi}$ itself, and therefore it appears not to be possible to extract non-symmetric contributions to $\tZ_{\phi}$ from the perturbative computation. However, in the main text we work backwards from Eq.~\eqref{gamredgen} to deduce the non-symmetric contribution in a general scheme, starting from the fact that the anomalous dimension in $\MSbar$ is symmetric and has no 1PR contributions.

\end{document}